\documentclass[journal,draftcls,onecolumn,12pt,twoside]{IEEEtran}

\usepackage{booktabs}
\usepackage{multicol}
\usepackage{multirow}
\normalsize
\usepackage{fancyhdr, epsfig, epsf, amsthm, amsmath, amssymb, amsfonts, subfigure,color,dsfont}
\usepackage{cite}
\usepackage{float}
\usepackage{textcomp}
\usepackage{xcolor}
\usepackage{dblfloatfix} 
\usepackage[most]{tcolorbox}
\tcbuselibrary{breakable} 
\usepackage{url}
\usepackage{hyperref}
\usepackage[keeplastbox]{flushend}
\usepackage{amsfonts}
\usepackage{amssymb}
\usepackage{amsmath}               
  {
      \theoremstyle{plain}
      
  }
\usepackage{algorithm}
\usepackage{algpseudocode}
\usepackage[utf8]{inputenc}

\usepackage{array}
\usepackage{booktabs}

\newtheorem{lemma}{Lemma}

\ifCLASSINFOpdf
\else
\fi

\DeclareMathOperator*{\argminA}{arg\,min} 

\begin{document}

\title{\fontsize{18}{22}\selectfont Predictive Control and Communication Co-Design via Two-Way Gaussian Process Regression and AoI-Aware Scheduling}

\author{Abanoub M.~Girgis,~\IEEEmembership{Student Member,~IEEE,}
        $^\dagger$Jihong~Park,~\IEEEmembership{Member,~IEEE,} 
        Mehdi~Bennis,~\IEEEmembership{Fellow,~IEEE}
        and~$^\ddagger$M\'erouane~Debbah,~\IEEEmembership{Fellow,~IEEE}
        
        \vspace{-1.5cm}
\thanks{A preliminary conference version of this work appeared in the proceedings of IEEE SPAWC-2020~\cite{9154304}.}
\thanks{A. Girgis and M. Bennis are with the Centre for Wireless Communications, University of Oulu, 90014 Oulu, Finland (e-mail: abanoub.pipaoy@oulu.fi, mehdi.bennis@oulu.fi).}
\thanks{$^\dagger$J. Park is with the School of Information Technology, Deakin University, Geelong, VIC 3220, Australia (e-mail: jihong.park@deakin.edu.au).}
\thanks{$^\ddagger$M\'erouane Debbah is with Université Paris-Saclay, CNRS, CentraleSupélec,
91190,  Gif-sur-Yvette, France (e-mail:
merouane.debbah@centralesupelec.fr) and the Lagrange Mathematical and
Computing Research Center, 75007, Paris, France.}
}

\maketitle

\begin{abstract}
This article studies the joint problem of uplink-downlink scheduling and power allocation for controlling a large number of actuators that upload their states to remote controllers and download control actions over wireless links. To overcome the lack of wireless resources, we propose a machine learning-based solution, where only a fraction of actuators is controlled, while the rest of the actuators are actuated by locally predicting the missing state and/or action information using the previous uplink and/or downlink receptions via a Gaussian process regression (GPR). This GPR prediction credibility is determined using the age-of-information (AoI) of the latest reception. Moreover, the successful reception is affected by the transmission power, mandating a co-design of the communication and control operations. To this end, we formulate a network-wide minimization problem of the average AoI and transmission power under communication reliability and control stability constraints. To solve the problem, we propose a dynamic control algorithm using the Lyapunov drift-plus-penalty optimization framework. Numerical results corroborate that the proposed algorithm can stably control $2$x more number of actuators compared to an event-triggered scheduling baseline with Kalman filtering and frequency division multiple access, which is $18$x larger than a round-robin scheduling baseline.

\end{abstract}


\IEEEpeerreviewmaketitle

\section{Introduction}    
\label{Sec1.1.Introduction}

 Ultra-reliable and low-latency communication (URLLC) is a key enabler for ensuring the stability of wirelessly networked control systems in real-time~\cite{park2017wireless,park2020extreme}. By physically decoupling sensors, actuators, and controllers, the control system can exploit the recent progress in the fifth generation (5G) connectivity~\cite{bennis2018ultrareliable}, machine learning and edge computing~\cite{park2019wireless}, thereby spearheading many emerging applications ranging from large-scale smart industrial internet of things (IIoT)~\cite{liu2019taming} to autonomous platooning~\cite{zeng2019joint}. The success of these application relies on addressing several fundamental challenges  emanating from unstable and intermittent wireless connectivity, which incurs distorted and delayed control information receptions, degrading control stability. Wireless resource allocation and scheduling are thus instrumental in not only improving communication efficiency but also in guaranteeing control stability. 
 
 
    Resource allocation and scheduling of control systems play a key role in the context of wireless networked control systems (WNCS). Round-robin scheduling, is a static scheduling in which each sensor/controller periodically transmits the state/action to a controller/actuator with a fixed transmission power and a predefined repeating order~\cite{hespanha2007survey,schenato2007foundations}. The scheduling method in~\cite{hespanha2007survey,schenato2007foundations} maintains stability for a small number of control systems, but fails to stabilize a large number of control systems since the scheduling decisions are not adaptive to the CSI, which hinder control performance. Dynamic round-robin scheduling in~\cite{xu2013stability,liu2003framework} is a channel-aware scheduling adapting the transmission power to the CSI to ensure reliability of the transmitted signal and minimize the system energy. However, it does not guarantee stability for a large number of control systems, since the waiting period of these control systems to update the state/action information is proportional to the number of served control systems. Moreover, the scheduling decisions are not adaptive to the control state that results in wasting wireless resources. To efficiently utilize communication resources, the event-triggered scheduling suggested in~\cite{cervin2008scheduling,postoyan2011event} is a dynamic scheduling method in which one control system with the largest state instability is scheduled at each time to transmit the state/action with fixed transmission power. The previously mentioned scheduling methods save wireless communication resources since the scheduling decisions are based on the control state but they are CSI-agnostic which may result in control performance degradation. In addition, these scheduling methods can not support a large number of control systems owing to the outdated control action dependent on the last received state.
    
    The control-aware communication scheduling suggested in~\cite{eisen2019control} incorporates in the scheduling decisions both control and channel states to minimize the overall transmission time of the scheduled control systems. Therein, the scheduling decisions are based on communication reliability that is sensitive to the control state and system dynamics. However, in the absence of wireless communication, the outdated applied actions over ideal channels are based on the last received state, hindering control stability. While the previously mentioned scheduling methods in~\cite{hespanha2007survey,schenato2007foundations,liu2003framework,xu2013stability,cervin2008scheduling} are effective in small-scale control environments, as pointed out in~\cite{zhao2019toward}, the communication and control designs in~\cite{hespanha2007survey,schenato2007foundations,liu2003framework,xu2013stability,cervin2008scheduling} are separated from each other, limiting their adoption for supporting a large number of control systems. This mandates a tight co-design of reliable communication and stable control operations for optimal resource allocation and scheduling.
    
   \subsection{Contributions}
    \label{sec1.2.Contributions}
 Spurred by the aforementioned motivation, in this work we consider a scenario consisting of a large number of control systems connected through wireless links. To overcome the lack of wireless resources, we propose a GPR based solution in which only a fraction of control systems is controlled by communicating with the controller, while the remaining control systems are controlled by locally predicting the missing state/action via GPR, using the previous received state/action. For each control system, a controller calculates the action using a linear quadratic regulator (LQR) whose input, i.e., either the estimated state using a minimum mean square error (MMSE) estimator if a sensor-controller pair of a control system is scheduled in the uplink (UL) or the predicted state using GPR at the controller when it is not scheduled. Then, to regulate the control system, an actuator applies either the estimated action  using an MMSE estimator if the controller-actuator pair of a control system is scheduled in the downlink (DL), or the predicted action using GPR at the actuator if it is not scheduled. The centralized scheduler shared among all control systems schedules at most one sensor-controller pair in the UL and one controller-actuator pair in the DL, based on both control and channel states. Here, control stability depends on the GPR prediction credibility~\cite{yan2009gaussian} that is determined by the age of information (AoI) of the latest received signal, and the reliability of the received signal that is dictated by transmission power, highlighting the importance of the communication and control co-design.  
 
 Given the proposed predictive control framework, we focus on the problem of jointly optimizing the UL-DL scheduling and power allocation so as to minimize the network-wide average AoI and transmission power while guaranteeing communication reliability and control stability. To solve the formulated non-convex stochastic optimization problem, we develop a dynamic control algorithm using Lyapunov optimization. Considering an inverted pendulum, numerical results demonstrate that the proposed scheduling method can stably support $2$x more control systems compared to the event-triggered scheduling with Kalman filtering and  frequency division multiple access (FDMA), which is $18$x larger than a time-triggered scheduling baseline. Furthermore, the results show that the proposed predictive control algorithm is more communication efficient while achieving a faster control stability than the time-triggered and event-triggered control baselines, highlighting the effectiveness of the UL-DL decoupled scheduling and the use of two-way GPRs at both controller and actuator sides.
 The remainder of this paper is organized as follows. In Section~\ref{Sec2.1.System Model}, we specify the WNCS architecture including the system models of the control, communication, and GPR based approach. In Section~\ref{sec3.Comm_Control_Codesign}, we formulate the communication, control, GPR based co-design optimization problem and propose the stability-aware scheduling algorithm by leveraging Lyapunov optimization framework to solve the co-design problem in Section~\ref{sec4.Lyapunov_Optimization}. In Section~\ref{Sec5.Simulation_Results} and Section~\ref{Conclusion}, we  present  simulations results and conclude the paper.
  

\begin{figure}[t]
\centering
\includegraphics[width = \columnwidth]{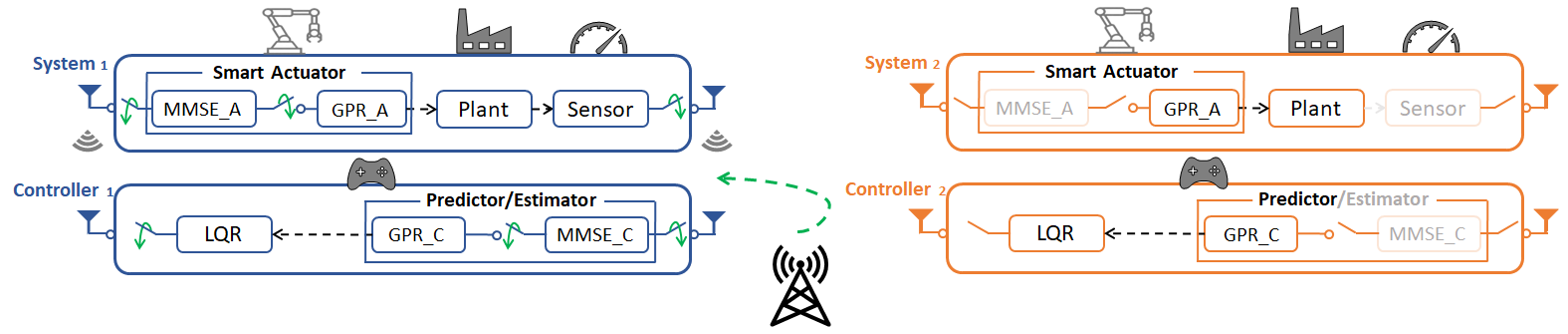}
\caption{ An illustration of $M=2$ WNCSs operated via both state/action measurement by remote sensors/controllers and state/actions prediction by GPR.}
\label{fig1}
\end{figure}

\section{System Model}
\label{Sec2.1.System Model}

 \subsection{Wireless Networked Control System Architecture} 
 \label{Sec2_1.Control_System_Arch}

 As depicted in Fig.~\ref{fig1}, the WNCS architecture under study consists of a set $\mathcal{M}$ of $M$ independent linear \emph{control systems} over a shared wireless channel. Each control system comprises a \emph{plant}, a \emph{sensor} that measures the plant's state, and \emph{actuator} that takes an action to control the plant's state. The action is computed by a remote \emph{controller} based on the control and channel states. To this end, the plant's state is received by the controller in the \emph{uplink}, and the controller's action is received by the actuator in the \emph{downlink}. To avoid interference under limited communication bandwidth, each UL or DL channel is allocated to a single sensor-controller pair or  controller-actuator pair per unit time, respectively. The rest of the  sensor-controller and controller-actuator pairs without receptions locally predict their missing control states and actions, respectively, based on their previously received information, to be detailed in Sec.~\ref{sec2_4.State_and_Action_Prediction}. To be specific, the plant's state of control system $i \in \mathcal{M}$ at discrete control time $k \in \mathbb{Z}_{+}$ is denoted by $\mathbf{x}^{u}_{i,k} \in \mathbb{R}^{D}$. For a received action at an actuator $\mathbf{u}^{a}_{i,k} \in \mathbb{R}^{P}$ based on the computed action at controller $\mathbf{u}^{d}_{i,k} \in \mathbb{R}^{P}$, the state evolution of control system $i$ at time $k$ is described by the discrete-time linear time-invariant (LTI) system as follows: \vspace{-3pt}
 { \small \begin{equation} 
 \label{eq1}
 \mathbf{x}^{u}_{i,k+1} = \mathbf{A}_{i} \mathbf{x}^{u}_{i,k} + \mathbf{B}_{i} \mathbf{u}^{a}_{i,k} + \mathbf{w}_{k}, 
 \end{equation} } \normalsize  where $\mathbf{A}_{i} \in \mathbb{R}^{D \times D}$ is a fixed state transition matrix of the $i$-th control system, $\mathbf{B}_{i} \in \mathbb{R}^{D \times P}$ is a fixed control action matrix of the $i$-th control system, and $\mathbf{w}_{k} \in \mathbb{R}^{D}$ is the plant noise at time $k$ which is independent and identically distributed (IID) Gaussian noise with zero mean and covariance matrix $\mathbf{W}$. Here, to avoid a non-trivial problem, $\mathbf{A}_{i}$ is assumed to be unstable, i.e., $\mathbf{A}_{i}$'s spectral radius $\rho(\mathbf{A}_{i}) = \text{max} \{ |\lambda_{1}(\mathbf{A}_{i})|, \cdots,| \lambda_{D}(\mathbf{A}_{i})| \} $ is larger than unity, where $\lambda_D(\mathbf{A}_{i})$ is the $D$-th eigenvalue of $\mathbf{A}_i$. This implies that the plant's state infinitely grows over time unless a proper control action $\mathbf{u}^{a}_{i,k}$ is provided. To stabilize such control system, each time $k$, the following four phase operations are considered. 

  \begin{enumerate}
  \item \textbf{Sensing and Uplink Transmission} (at a sensor): A centralized scheduler located at the base station (BS) shared among all control systems decides which sensor-controller pair is scheduled to transmit and close its sensing loop based on both the channel and control states. Then, the scheduled sensor transmits its state to its controller over a wireless UL fading channel using analog uncoded communication to be elaborated in Sec.~\ref{sec2_2.State_and_Action_Communications}.
  
  \item \textbf{State Reception or Prediction} (at a controller): If the sensor-controller pair is scheduled, the controller obtains the current estimated state using MMSE estimator, and predicts the next state via GPR to be discussed in Sec.~\ref{sec2_4.State_and_Action_Prediction}. Otherwise, the controller directly predicts the current and the next states based on the state history using GPR. The current predicted state by GPR is fed to the LQR to calculate the action unless the estimated state by the MMSE estimator is provided. The future predicted state is fed to the centralized scheduler to make the scheduling decisions.
  
 \item \textbf{Action Computation and Downlink Transmission} (at a controller): For a given control state, the controller computes the optimal action using LQR~\cite{aastrom2013adaptive}. The controller transmits the computed action to the scheduled actuator over a wireless DL fading channel using analog uncoded communication to be elaborated in Sec.~\ref{sec2_2.State_and_Action_Communications}.
  
  \item \textbf{Action Reception or Prediction} (at an actuator): If the controller-actuator pair is scheduled, the actuator obtains the current estimated action using a MMSE estimator, and predicts the next action via an action GPR to be discussed in Sec.~\ref{sec2_4.State_and_Action_Prediction}. Otherwise, the actuator directly predicts the current and next actions based on the action history using GPR. For a given action, the actuator takes an action and subsequently the plant's state is updated according to the dynamics in \eqref{eq1}.
 \end{enumerate}

 \subsection{State and Action Communications}
 \label{sec2_2.State_and_Action_Communications} 

 The UL state and DL action communications are elaborated, in terms of the received signal, signal-to-noise ratio (SNR), scheduling, and age-of-information (AoI) as follows.

 \noindent\textbf{Noisy State and Action Receptions}:\quad At the control time slot $k$, the received signal $\mathbf{y}^{l}_{i,k}$ at the $l$-th communication at the receiver of control system $i$ is represented as 
{\small \begin{align} 
 \label{eq2}
 \mathbf{y}^{l}_{i,k}  = \sqrt{P^{l}_{i,k}} \mathbf{C}_{i} \mathbf{H}^{l}_{i,k}  \mathbf{q}^{l}_{i,k} + \mathbf{n}^{l}_{k},
 \end{align} } \normalsize where $l \in \{ u, d \}$ represents a communication indicator between the transmitter-receiver pair in which $l = u$ refers to the UL state communication between the sensor-controller pair while $l = d$ refers to the DL action communication between the controller-actuator pair. The transmitted signal, in the UL state communication (i.e., $\mathbf{q} = \mathbf{x}$ and $l = u$), $\mathbf{x}^{u}_{i,k} = [x^{u}_{i,k}(1) \cdots x^{u}_{i,k}(D)]$  is the plant's state transmitted by the sensor of a control system $i$ at time $k$ such that $\mathbb{E} \{ |x^{u}_{i,k} (\iota)|^{2} \} = 1, \forall \iota \in \{1,\cdots,D \}$. The transmitted signal, in the DL action communication (i.e., $\mathbf{q} = \mathbf{u}$,  $l = d$), $\mathbf{u}^{d}_{i,k} = [u^{d}_{i,k}(1) \cdots u^{d}_{i,k}(P)]$ is the action transmitted by the controller to an actuator of a control system $i$ at time $k$ such that $\mathbb{E} \{ |u^{d}_{i,k} (p)|^{2} \} = 1, \forall p \in \{1,\cdots,P \}$. The matrix $\mathbf{H}^{l}_{i,k} \in \mathbb{R}^{\mathcal{F} \times \mathcal{F} } $  represents the wireless channel of the $l$-th communication between the transmitter-receiver pair of a control system $i$ at time $k$, and $\mathcal{F} \in \{ D, P \}$ represents the dimensions of the transmitted state or action, respectively. The channel is modeled as a Rayleigh block fading which is static and flat-fading within either UL or DL transmission time. Channel statistics are perfectly known at the transmitters and receivers, and $P^{l}_{i,k} \in \left[ 0,P^{l}_{max} \right]$ is the transmission power of control system $i$ at time $k$ with total transmission power $P^{l}_{max}$. Lastly, $\mathbf{n}^{l}_{k}$ is the additive white Gaussian noise at the receiver with zero-mean and covariance matrix $ \mathbb{E}\{ \mathbf{n}_{k}^{l^{T}} \mathbf{n}^{l}_{k} \} = N_{0} \mathbf{I}_{\mathcal{F}}$, where $N_{0}$ is the measurement noise variance, and $\mathbf{I}_{\mathcal{F}}$ is the $\mathcal{F} \times \mathcal{F}$ identity matrix. The matrix $\mathbf{C}_{i} \in \mathbb{R}^{\mathcal{F} \times \mathcal{F}}$ is the observation matrix of the control system $i$ that equals the identity matrix in the DL action communication and equals the identity matrix in the UL state communication to characterize the full-state observations. The measurement noise $\mathbf{n}^{l}_{k}$ and the plant's noise $\mathbf{w}_k$ in \eqref{eq1} are uncorrelated zero-mean Gaussian noise with unit variance. Hence, the SNR at the receiver of the $l$-th communication of a control system $i$ at time $k$ is given as \vspace{-2pt} {\small 
 \begin{align} 
 \label{eq3}
 \text{SNR}^{l}_{i,k} = \frac{P^{l}_{i,k} \Vert \mathbf{H}^{l}_{i,k} \Vert^{2}}{N_{0}},
 \end{align} } where SNR in~\eqref{eq3} is equivalent to the signal-to-distortion ratio (SDR) as a result of the channel theoretical limit in which the rate-distortion function of the source equals to the channel capacity and the number of source samples is matched to the number of channels under analog uncoded communications~\cite{hassanin2013analog}. In this work, we consider analog uncoded communications, in which the discrete-time continuous amplitude source samples are amplified and transmitted to a receiver over wireless channel. Compared to digital communications, analog communications are favorable for achieving low latency thanks to skipping encoding and decoding operations, at the cost of signal distortion during channel propagation~\cite{hassanin2013analog}. Furthermore, analog uncoded communication is robust to channel conditions and performs well at different SNRs compared to digital communication that is sensitive to any degradation in channel conditions. To ensure reliable communication for control stability, the successful decoding of the transmitted signal is described by the indicator function $ \mathbb{I}_{\{  \text{SNR}^{l}_{i,k} \geq \text{SNR}^{l}_{\text{th}} \} }$ for a target $\text{SNR}$ threshold $\text{SNR}^{l}_{\text{th}}$. 
 
 \noindent\textbf{Scheduling and AoI}:\quad
 At  time $k$, the centralized scheduler located at the BS and shared among all control systems schedules at most one sensor-controller pair of  control system $i$ in the UL state communications, and at most one controller-actuator pair of the control system $i$ in the DL action communications. Let $\alpha^{l}_{i,k} \in \{0,1 \}$ be the scheduling variable of the $l$-th communication of the control system $i$ at time $k$, where $\alpha^{l}_{i,k}  = 1$ when the transmitter-receiver pair of the $l$-th communication of control system $i$ is scheduled at time $k$ and $\alpha^{l}_{i,k} = 0$ otherwise. 

The freshness of the received information is measured using AoI, i.e., the number of elapsed time since the generation of the latest received information~\cite{kosta2017age}.  AoI is composed of the inter-arrival time that is defined as the time elapsed between two consecutive update generations and the service time defined as the transmission time of update information. In analog uncoded communications,  AoI depends only on the inter-arrival time since the service time is deterministic based on channel bandwidth. Hence, the AoI of the $l$-th communication of the control system $i$ at the receiver linearly increases with time if it is not scheduled or its SNR is below a threshold. Formally, the AoI of the $l$-th communication of control system $i$ at the receiver is: \vspace{-2pt}
 {\small \begin{align} 
 \label{eq4}
 \beta^{l}_{i,k} = \left\{ \begin{array}{ll}  1 + \beta^{l}_{i,k-1}, &\text{if} \; \xi^{l}_{i,k} = 0, \\  1, & \text{o.w.s.},
 \end{array}
 \right.
 \end{align} } where $\beta^{l}_{i,k} \in \mathbb{Z}_{++}$ is the AoI of the $l$-th communication of the control system $i$ at time $k$ at the receiver, and $\xi^{l}_{i,k} = \alpha^{l}_{i,k} \, \mathbb{I}_{ \{ \text{SNR}^{l}_{i,k} \geq \text{SNR}^{l}_{th}\}}$ is the transmission indicator variable of  $l$-th communication of control system $i$ at time $k$ that depends on both  scheduling variable and  SNR indicator function.

\subsection{State and Action Estimation Over Noisy Communication Links}
 \label{sec2_3.State_and_Action_Estimation} The UL received states and DL  received actions are distorted by Rayleigh fading channels. The original signals are estimated using the MMSE estimator as detailed next. When one transmitter-receiver pair of the $l$-th communication of  control system $i$ is scheduled, i.e., $\alpha_{i,k}^{l}=1$, the receiver applies the MMSE estimator to restore the original signal from the noisy received signal in \eqref{eq2}. The resultant estimated signal $\bar{\mathbf{q}}^{l}_{i,k}$ is given~as: 
 {\small \begin{align} 
 \label{eq5}
 \bar{\mathbf{q}}^{l}_{i,k}=\mathbb{E} \{ \mathbf{q}^{l}_{i,k} \vert \mathbf{y}^{l}_{i,k} \} = \mathbf{G}^{l}_{i,k} \mathbf{y}^{l}_{i,k}   = \mathbf{q}^{l}_{i,k} + \mathbf{v}^{l}_{i,k},  
 \end{align} }where $\mathbf{G}^{l}_{i,k} \in \mathbb{R}^{\mathcal{F} \times \mathcal{F}}$ is the linear MMSE matrix at the receiver of the $l$-th communication of the control system $i$ at time $k$ that minimizes the mean-squared error (MSE) between the original and estimated signals as  \vspace{-2pt}
 {\small \begin{align}
 \label{eq5+}
 \mathbf{G}^{l}_{i,k} = \sqrt{P^{l}_{i,k}} \mathcal{S}_{q} \mathbf{H}^{l^{T}}_{i,k} \left( P^{l}_{i,k} \mathbf{H}^{l}_{i,k} \mathcal{S}_{q} \mathbf{H}^{l^{T}}_{i,k} + N_{0} \mathbf{I}_{\mathcal{F}} \right)^{-1}
 \end{align}} The term $\mathbf{v}^{l}_{i,k}$ in~\eqref{eq5} is the MMSE estimation error following a zero-mean Gaussian random vector with the covariance matrix $\mathbf{V}^{l}_{i,k} \in \mathbb{R}^{\mathcal{F} \times \mathcal{F}}$. Following~\cite{oppenheim2010introduction}, we assume that $\mathbf{q}^{l}_{i,k}$ follows a zero-mean Gaussian distribution with the covariance matrix $\mathcal{S}_{q} \in \mathbb{R}^{\mathcal{F} \times \mathcal{F}}$, then we have \vspace{-3pt}  { \small \begin{align}
 \label{eq6}
 \mathbf{V}^{l}_{i,k} &=\mathbb{E} \{ \mathbf{v}^{l}_{i,k} \mathbf{v}_{i,k}^{l^{T}} \} = \mathbb{E} \Big\{ \left( \bar{\mathbf{q}}^{l}_{i,k}   - \mathbf{q}^{l}_{i,k} \right)\left(\bar{\mathbf{q}}^{l}_{i,k}   - \mathbf{q}^{l}_{i,k}\right)^{T}  \Big\} = \mathcal{S}_{q} - \mathbf{G}^{l}_{i,k} \sqrt{P^{l}_{i,k}} \mathbf{H}^{l}_{i,k}  \mathcal{S}_{q}
 \end{align} }

 \subsection{State and Action Prediction Without Communication}
 \label{sec2_4.State_and_Action_Prediction} When one transmitter-receiver pair of the $l$-th communication of the control system $i$ at time $k$ is not scheduled, i.e., $\alpha^{l}_{i,k} = 0$, a receiver applies a number of parallel GPRs proportional to the missing signal dimensions to predict both the missing current signal and the next signal using the previously received signals. Each individual GPR learns the functional relationship $g \in \mathbb{R}$ between the control discrete-time $ k' \in \mathbb{Z}_{+}$ and each output of the received signal. This means that each output of the MMSE estimated signal $\bar{\mathbf{q}}^{l}_{i,k'}$ in \eqref{eq5} is the state observation $\bar{\mathbf{x}}^{u}_{i,k'} \in \mathbb{R}^{D}$ in the UL and the action $\bar{\mathbf{u}}^{d}_{i,k'} \in \mathbb{R}^{P}$  in the DL. This is accomplished by learning a latent function of the following regression model $\bar{q}^{l}_{i,k'}(j) = g_{j}(k') + \epsilon, \; j \in \{ 1, \cdots, \mathcal{F} \}, \; \forall i, \, l, \, k'$, where $\bar{q}^{l}_{i,k'}(j) \in \mathbb{R}$ is the $j$-th output of the estimated signal of the $l$-th communication of the control system $i$ at time $k'$, $g_{j}$ is a $j$-th output latent function, and $\epsilon \sim \mathcal{N} \left( 0, \sigma^{2}_{n} \right) $ is an IID Gaussian noise distribution with zero mean and variance $\sigma^{2}_{n}$ that accounts for the measurements or modeling errors~\cite{alvarez2012kernels}. Specifically, to predict the missing received signal $\bar{\mathbf{q}}^{l}_{i,k}$ of the $l$-th communication of the control system $i$ at test time $k$, we exploit $\mathcal{F}$ individual GPRs, where $\mathcal{F}$ represents the MMSE estimated signal dimensions, and feeding each individual GPR with a training set $\mathcal{D}^{l,j}_{i,n_{l}}$ of each output of the previous received signals associated with its observation time $k'$, given as $\mathcal{D}^{l,j}_{i,n_{l}} = \{( k', \xi^{l}_{i,k'} \, \bar{q}^{l}_{i,k'}(j) ) \vert \, j = 1, \cdots, \mathcal{F},\; k' = 1, \cdots, n_{l},\; i=1,\cdots,M, \;  l \in \{ u,d \} \}$. Here, $n_{l} = \sum_{k'} \xi^{l}_{i,k'}$ counts the number of received signals of the $l$-th communication until time $k'$ in the training set of the control system $i$. Hence, the last time instant in which the transmitter of the $l$-th communication of the control system transmitted its observation to the receiver is given as $\tilde{n}_{l} = k' - \beta^{l}_{i,k'} + 1 $. It is obvious that a large value of AoI decreases the number of observations at the receiver that affects the signal prediction credibility at a particular level.

 In each individual GPR, according to the Gaussian process (GP) characteristics where any finite subset of random variables taken from a realization of a GP follows a joint Gaussian distribution, each $j$-th output latent function $g_{j}$ of the vector-valued latent function $\mathbf{g}(k) = \left[ g_{1}(k) \cdots g_{\mathcal{F}}(k)\right]$ is assumed to follow a GP as $ g_{j} \left(k \right) \sim \mathcal{GP} \left( m_{j} \left(k \right), \mathcal{R}_{j}(k,k') \right)$, where $m_{j} \left( k \right)$ is the mean function of the $j$-th output of the missing received signal which is usually taken as zero without loss of generality~\cite{rasmussen2003gaussian}, and $\mathcal{R}_{j}(k,k')$ is the covariance function of the $j$-th output of the missing received signal between the outputs at time $k$ and $k'$ that defines the correlation between the outputs according to the input times. It is noted that the stationary covariance function between the outputs is based on the difference between their corresponding input times $|k - k'|$ in which the two outputs are strongly correlated if their corresponding input times are sufficiently close to each other. Since we focus on time-series data, we utilize information from previously received signals to describe the current data depending on the past observations. Hence, we use a squared exponential kernel function coupled with a periodic kernel function, to model the correlation between the outputs according to their temporal behaviours, as defined in~\cite{rasmussen2003gaussian} \vspace{-2pt} { \small \begin{align}  
 \label{eq7}
 \mathcal{R}(k,k') = h_{q}^{2} \exp \left[ \frac{-\left( k - k'\right)^{2}}{2h^{2}_{k}}  \right] + \exp \left\lbrace -2 \sin^{2} \left[ \nu \pi \left(k - k' \right) \right] \right\rbrace,
 \end{align}} where the first term represents the stationary covariance function that depends on when the signal $\vert k - k' \vert$  was received with $h_{k}$ and $h_{q}$ being hyperparameters representing the time-scaling and output-scaling of a squared exponential function, respectively, and the second term gives the periodicity with hyperparameter $\nu$  representing  frequency. For a set of j-th output observations $\bar{\mathbf{q}}^{l}_{i}(j) = \{ \bar{q}^{l}_{i,1}(j), \cdots, \bar{q}^{l}_{i,n_{l}}(j) \}^{T} $ and the associated observation times $\mathbf{k}' = \{ 1, \cdots, n_{l} \}^{T}$, the joint distribution of the $j$-th output past observations $\bar{\mathbf{q}}^{l}_{i,k'}(j)$  together with the $j$-th output $g_{j}(k)$ at  test time $k$ is given as \vspace{-2pt} {\small \begin{align}
 \label{eq7+}
 \left[ \begin{array}{c}
     \bar{\mathbf{q}}^{l}_{i}(j)  \\
     g_{j}(k)
 \end{array}  \right]  \sim \mathcal{N} \left( \left[ \begin{array}{c}
     \mathbf{0}  \\
      0
 \end{array}  \right] , \left[ \begin{array}{cc}
     \mathbf{R}_{j}(\mathbf{k}',\mathbf{k}') &  \mathbf{r}_{j}(\mathbf{k}',k) \\
     \mathbf{r}_{j}(k,\mathbf{k}') & \mathcal{R}_{j}(k,k)
 \end{array}  \right]  \right),
 \end{align}} where $\mathcal{R}_{j}\left(k,k \right) \in \mathbb{R}$ is the prior covariance function of $j$-th output observation at a test time $k$, and  $\mathbf{R}_{j}(\mathbf{k}',\mathbf{k}') \in \mathbb{R}^{n_{l} \times n_{l}} $ is the symmetric and positive semi-definite covariance matrix of $j$-th output past observations with the elements $\mathcal{R}_{j} \left( {\mathbf{k}'(a),\mathbf{k}'(b) } \right)$ for $a,b = 1, \cdots, n_{l}$. Following \cite{alvarez2012kernels}, we treat the prediction mean as the $j$-th output predicted signal $\hat{q}^{l}_{i,k}(j)$, the posterior distribution of $g_{j}\left(k\right)$ at test time $k$ based on the training set $\mathcal{D}^{l,j}_{i,n_{l}}$ can be analytically derived as \vspace{-2pt}  { \small
 \begin{align}  
 \label{eq8}
 \text{Pr}\left( g_{j} \left(k \right) \vert   \mathcal{D}^{l,j}_{i,n_{l}}, k, \mathbf{\Theta}_{j} \right) &\sim \mathcal{N} \left( \hat{q}^{l}_{i,k}(j), \sigma^{2}_{i,k}(j)  \right).
 \end{align} }
 Following~\cite{alvarez2012kernels}, the $j$-th output prediction mean $\hat{q}^{l}_{i,k}\left(j \right)$, and the $j$-th output prediction variance $\sigma^{2}_{i,k}(j)$ are respectively given as\vspace{-2pt}{ \small  \begin{align}
 \hat{q}^{l}_{i,k}(j) &= \mathbf{r}_{j}(k,\mathbf{k}')^{T}  \mathbf{R}_{j}(\mathbf{k}',\mathbf{k}')^{-1} \bar{\mathbf{q}}^{l}_{i}(j) = q^{l}_{i,k}(j) + e^{l}_{i,k}(j), \label{eq9a}\\
 \sigma^{2}_{i,k} (j) & = \mathbb{E} \Big\{ \left(  \hat{q}^{l}_{i,k}(j) -  q^{l}_{i,k}(j)   \right) \left( \hat{q}^{l}_{i,k}(j)  - q^{l}_{i,k}(j)  \right)^{T} \Big\} = \mathbb{E} \{e^{l}_{i,k}(j) e^{l^{T}}_{i,k}(j)\} \label{eq9b}\\ & = \mathcal{R}_{j}(k,k) - \mathbf{r}_{j}(k,\mathbf{k}')\mathbf{R}_{j}(\mathbf{k}',\mathbf{k}')^{-1} \mathbf{r}_{j}(k,\mathbf{k}')^{T}, \nonumber
 \end{align} }
 where $\mathbf{r}_{j}(\mathbf{k}',k) \in \mathbb{R}^{n_{l} \times 1}$ is $j$-th output observation covariance between the outputs at the $n_{l}$ observation times and a test time $k$, and the term $e^{l}_{i,k}(j)$ is $j$-th output prediction error defined as the difference between true and predicted outputs. Moreover, $\mathbf{\Theta}_{j}$ in~\eqref{eq8} is the $j$-th output hyperparameters of the covariance function $\mathcal{R}$. Finally, the predicted signal at the receiver of the $l$-th communication of control system $i$ at time $k$ and its prediction error covariance matrix are \vspace{-2pt}  { \small \begin{align}
 \label{eq9+a}
 \hat{\mathbf{q}}^{l}_{i,k} & = \{ \hat{q}^{l}_{i,k}(1) \cdots \hat{q}^{l}_{i,k}(\mathcal{F}) \}^{T} = \mathbf{q}^{l}_{i,k} + \mathbf{e}^{l}_{i,k}, \\
 \label{eq9+b}
\mathbf{\mathcal{J}}^{l}_{i,k} & = \left[ \begin{array}{ccc}
    \sigma^{2}_{i,k}(1) &  \cdots   & 0   \\
                \vdots  &  \ddots   & \vdots  \\
                0       &  \cdots   & \sigma^{2}_{i,k}(\mathcal{F})
\end{array} \right].
\end{align} }
\subsection{Action Computation and Actuation}
\label{Sec:Sys_Actuation}

By feeding the estimated or predicted state, the controller computes the action using LQR. Then, the actuator applies the estimated or predicted action to stabilize the state as detailed next.

 \noindent\textbf{Action Computation After State Estimation/Prediction:}\quad 
 For a given estimated state (i.e., MMSE output) in~\eqref{eq5} or predicted state (i.e., GPR output) in~\eqref{eq9+a}, at time $k$, the state $\mathbf{x}_{i,k}^{c}$ available at the controller based on the UL transmission indicator variable is given as { \small \begin{align}
 \label{eq20}
 \mathbf{x}_{i,k}^{c} = \xi^{u}_{i,k} \bar{\mathbf{x}}^{u}_{i,k} + \left( 1 - \xi^{u}_{i,k}  \right) \hat{\mathbf{x}}^{u}_{i,k}.
 \end{align} }
 The received state $\mathbf{x}_{i,k}^{c}$ is used in the LQR located at the controller, and the optimal action of a control system $i$ at time $k$ is given by the following linear feedback control law as \vspace{-3pt} { \small
\begin{align}
\mathbf{u}^{d}_{i,k} = - \mathbf{\Phi}_{i}  \mathbf{x}^{c}_{i,k}, \label{eq11}
\end{align} } where $\mathbf{u}^{d}_{i,k} \in \mathbb{R}^{P}$ is the computed action at the controller, $\mathbf{\Phi}_{i} = \left( \mathbf{Z}^{u} + \mathbf{B}^{T}_{i} \mathbf{P} \mathbf{B}_{i} \right)^{-1}  \mathbf{B}_{i}^{T} \mathbf{P}\mathbf{A}_{i}$ is the feedback gain matrix of the control system $i$, $\mathbf{Z}^{s} \in \mathbb{S}_{+}^{D \times D}$ is a positive semi-definite weight matrix of the state deviation cost, and $\mathbf{Z}^{u} \in \mathbb{S}_{++}^{P \times P}$ is a positive definite weight matrix of the action cost. The term  $\mathbf{P} = \mathbf{A}_{i}^{T} \mathbf{P} \mathbf{A}_{i} - \mathbf{A}_{i}^{T} \mathbf{P} \mathbf{B}_{i} \left( \mathbf{B}^{T}_{i} \mathbf{P} \mathbf{B}_{i} + \mathbf{Z}^{u} \right)^{-1} \mathbf{B}_{i}^{T} \mathbf{P} \mathbf{A}_{i} + \mathbf{Z}^{s}$ is the unique positive definite matrix which satisfies the discrete-time algebraic Riccati equation (DARE). Then, the controller transmits the computed action $\mathbf{u}^{d}_{i,k}$ in \eqref{eq11} to an actuator of control system $i$ at time $k$ in the DL, if $\xi_{i,k}^{d}=1$, as discussed in Sec.~\ref{sec2_2.State_and_Action_Communications}.

\noindent\textbf{Actuation After Action Estimation/Prediction:}\quad
For a given estimated control action (i.e., MMSE output) in~\eqref{eq5} or predicted action (i.e., GPR output) in~\eqref{eq9+a}, at time $k$, the action $\mathbf{u}_{i,k}^{a}$ available at the actuator based on the DL transmission indicator variable is given as { \small
\begin{align}
\label{eq21}
\mathbf{u}_{i,k}^{a} = \xi^{d}_{i,k} \, \bar{\mathbf{u}}^{d}_{i,k} + \left( 1 - \xi^{d}_{i,k}  \right) \hat{\mathbf{u}}^{d}_{i,k}.
\end{align} } 
Note that the UL and DL transmission indicator variables are periodically generated by the centralized scheduler in which, within each unit control time duration, the UL state communication can be firstly activated for sensing the plant's state based on the UL transmission indicator variable. Then, the DL action communication can be activated for actuation based on the DL transmission indicator variable. Consequently, for a given pair of UL and DL transmission indicator variables with~\eqref{eq20} and~\eqref{eq21}, the actuator takes a control action that changes the plant's state of the control system $i$ at time $k$ in~\eqref{eq1} into four cases of state evolution as follows: \vspace{-5pt} { \small
 \begin{align}
   \mathbf{x}^{o}_{i,k+1}  =  \mathbf{A}_{i} \mathbf{x}^{u}_{i,k} - \mathbf{B}_{i} ( \mathbf{\Phi}_{i} \hat{\mathbf{x}}^{u}_{i,k} +  \mathbf{e}^{d}_{i,k} )  + \mathbf{w}_{k}, \; &\text{if} \; \xi^{u}_{i,k} = 0 \;  \text{and} \; \xi^{d}_{i,k} = 0, \; \text{ (open-loop)},\label{eq22.1} \\ \mathbf{x}^{s}_{i,k+1} = \mathbf{A}_{i} \mathbf{x}^{u}_{i,k} - \mathbf{B}_{i} ( \mathbf{\Phi}_{i} \bar{\mathbf{x}}^{u}_{i,k} +  \mathbf{e}^{d}_{i,k} )  + \mathbf{w}_{k},\; &\text{if} \; \xi^{u}_{i,k} = 1 \;  \text{and} \; \xi^{d}_{i,k} = 0, \; \text{ (sensing-loop)},\label{eq22.2} \\ \mathbf{x}^{a}_{i,k+1} =   \mathbf{A}_{i} \mathbf{x}^{u}_{i,k} - \mathbf{B}_{i} ( \mathbf{\Phi}_{i} \hat{\mathbf{x}}^{u}_{i,k} +  \mathbf{v}^{d}_{i,k} )  + \mathbf{w}_{k},\;  &\text{if} \; \xi^{u}_{i,k} = 0 \; \text{and} \; \xi^{d}_{i,k} = 1,\; \text{ (actuating-loop)},\label{eq22.3}  \\ \mathbf{x}^{c}_{i,k+1} = \mathbf{A}_{i} \mathbf{x}^{u}_{i,k} - \mathbf{B}_{i} ( \mathbf{\Phi}_{i} \bar{\mathbf{x}}^{u}_{i,k} +  \mathbf{v}^{d}_{i,k} )  + \mathbf{w}_{k}, \; &\text{if} \; \xi^{u}_{i,k} = 1 \;  \text{and} \; \xi^{d}_{i,k} = 1, \; \text{ (closed-loop)}.\label{eq22.4}
 \end{align} }
 
 \begin{figure}[t]
\centering
\includegraphics[trim=20 20 8 8,clip,width=.70\linewidth]{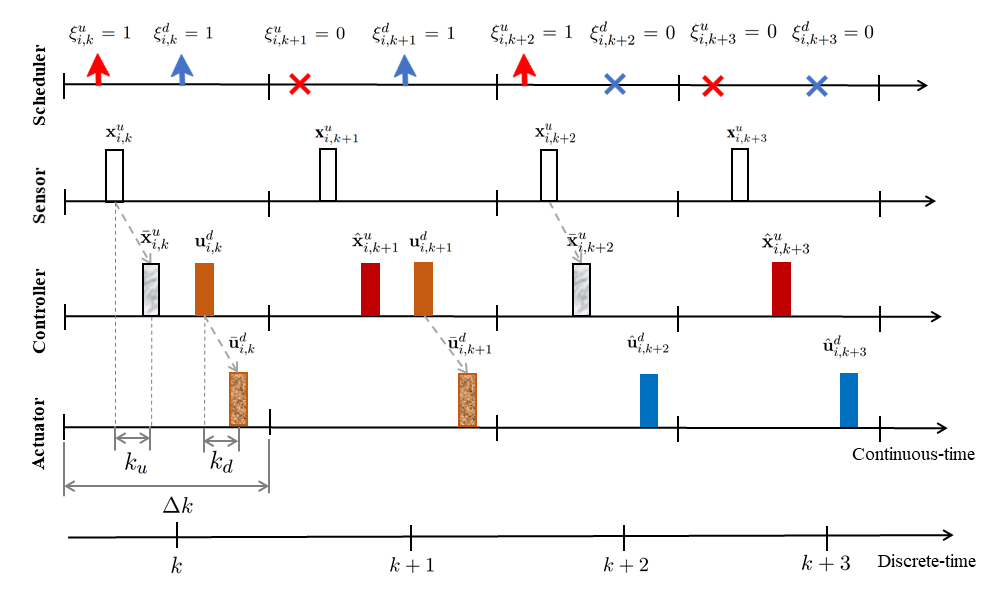}
\caption{ Timing diagram of the control system. First diagram illustrates the UL and DL transmission indicator variables generated by scheduler, second diagram illustrates uniform sampling by a sensor, third diagram illustrates the received/predicted state and calculated action at a controller. Fourth diagram illustrates the received/predicted action at an actuator.} 
\label{figtime}
\end{figure} \vspace{-25pt}
  \noindent\textbf{Timing diagram:}\quad Based on the UL and DL transmission indicator variables within each unit control time duration, the timing diagram of a control system is illustrated in Fig.~\ref{figtime}. The centralized scheduler shared among all control systems, within each unit control time duration, primarily transmits the UL and DL transmission indicator variables to the sensor, controller, and actuator nodes of all control systems. Then, the state is only transmitted by the sensor if the control system has a reliable UL communication and has valuable information affecting the control stability (i.e, $\xi^{u}_{i,k} = 1$) which result in saving wireless communication resources. After that, LQR located at the controller computes the action based on the state available at the controller in~\eqref{eq20} in which the predicted state, if $\xi^{u}_{i,k} = 0$, is applied to the LQR. Lastly, the action is transmitted by the controller if has a reliable DL communication and valuable information affecting the control stability (i.e, $\xi^{d}_{i,k} = 1$). This is a result of assuming that the UL and DL transmission indicator variables as being periodically transmitted by the centralized scheduler every control time duration, the controller periodically calculates the action depending on the available state, and the actuator periodically applies the action depending on the available action. Moreover the discrete-time control time $k$  equals the continuous-time control time duration unit $\Delta k$ comprising the UL and DL transmission times while ignoring the computational delay. 

\section{Communication Control Co-design}  
\label{sec3.Comm_Control_Codesign}

\subsection{Control-constrained Problem Formulation}
\label{sec3_1.Problem_Formulation}

Our objective is to minimize the total communication cost per control system subject to ensuring communication reliability and control stability. The total communication cost incorporates the AoI and transmission power since the AoI indirectly affects the control stability through the GPR prediction stability and  wireless resources consumption, while  transmit power affects communication reliability and  energy consumption. Formally speaking, we have:\vspace{-3pt} { \small
\begin{align}
\label{eq34} 
\mathcal{C} \left( \{ \bar{\beta}^{l}_{i} \}, \{ \bar{\hat{P}}^{l}_{i} \} \right) = \omega_{\beta_{l}} \sum_{i=1}^{M} \mathcal{G}_{\beta}( \bar{\beta}^{l}_{i})  +    \omega_{P_{l}} \sum_{i=1}^{M} \mathcal{G}_{P}( \bar{\hat{P}}^{l}_{i}), \qquad  \forall l \in \{ u,d \},
\end{align} } where the non-decreasing concave functions $\mathcal{G}_{\beta}(\beta) = \log (1+\beta)$  and  $\mathcal{G}_{P}(\hat{P}) = \log (1+\hat{P})$ are proportionally fair cost functions of the AoI and the transmission power function for each control system, respectively~\cite{li2008proportional}. The transmission power function that depends on the scheduling variable is given as $\hat{P}^{l}_{i,k} = \alpha^{l}_{i,k} P^{l}_{i,k}$, and the given positive weights $\omega_{\beta_{l}}$ and $\omega_{P_{l}}$ adjust the relative importance of the corresponding cost functions. Throughout this work, the following notation for the long-term time-averaged of any quantity $z$ is defined as $ \bar{z} \triangleq \underset{K\to \infty}{\lim\sup}  \frac{1}{K} \sum_{k=1}^{K} z$. In particular, $\bar{\beta}^{l}_{i}$ and $\bar{\hat{P}}^{l}_{i}$ are the long-term time-averaged of $\beta^{l}_{i}$ and $\hat{P}^{l}_{i}$, respectively.


 To evaluate control stability, we consider the quadratic Lyapunov function that measures the performance of each control system as a function of the state expressed as \vspace{-5pt} { \small \begin{align}
 \label{Lyapunov_fun}    
 \mathcal{L}(\mathbf{x}^{u}_{i,k}) = \mathbf{x}^{u^{T}}_{i,k} \, \mathcal{Z} \, \mathbf{x}^{u}_{i,k},   \qquad \forall \mathcal{Z} \in  \mathbb{S}^{D}_{++},
 \end{align}} where $\mathcal{Z} \in \mathbb{S}^{D}_{++}$ is a unique positive definite solution to the discrete Lyapunov equation $\mathbf{A}^{c^{T}}_{i} \mathcal{Z} + \mathcal{Z} \mathbf{A}^{c}_{i} = - \mathbb{I}_{D}$, and $\mathbf{A}^{c}_{i}$ is a closed-loop state transition matrix defined as $\mathbf{A}^{c}_{i} = \mathbf{A}_{i} - \mathbf{B}_{i} \mathbf{\Phi}_{i} $. Because the centralized scheduler has only access to the predicted state, the expected current value of $\mathcal{L}(\mathbf{x}^{u}_{i,k})$  is calculated in the following lemma.
 \begin{lemma}
 \label{lemma_1}
 Given the predicted state $\hat{\mathbf{x}}^{u}_{i,k}$ and the state prediction error covariance matrix $\mathcal{J}^{u}_{i,k}$ at the controller, the expected current value of $\mathcal{L}(\mathbf{x}^{u}_{i,k})$ is given as \vspace{-5pt}  { \small 
 \begin{align}
 \label{Lyapunov_Expec} 
 \mathbb{E}\left[ \mathcal{L}(\mathbf{x}^{u}_{i,k}) \vert \hat{\mathbf{x}}^{u}_{i,k} \right]  = \Vert \hat{\mathbf{x}}^{u}_{i,k} \Vert_{\mathcal{Z}^{\frac{1}{2}}}^{2} + \text{Tr} \left[  \mathcal{Z} \, \mathcal{J}^{u}_{i,k} \right], \qquad \forall \; \mathcal{Z} \in \mathbb{S}^{D }_{++}.
 \end{align} } \vspace{-30pt} \begin{proof} 
 Please refer to Appendix.\ref{App.lemma1} 
 \end{proof}
 \end{lemma}  Note that the expected current value of $\mathcal{L}(\mathbf{x}^{u}_{i,k})$  of the control system $i$ at time $k$ naturally grows as the predicted state and the prediction error get larger as a result of increasing AoI and/or the insufficiency of received observations number in the training set. The requirement of control stability is that the expected future value of $\mathcal{L}(\mathbf{x}^{u}_{i,k+1})$  should decrease at a given rate $\zeta_{i} \in (0 , 1 ]$ of its expected current value of $\mathcal{L}(\mathbf{x}^{u}_{i,k})$, which means the state of the control system is monotonically decreasing along trajectories, as \vspace{-5pt} { \small
 \begin{align}
 \label{Future_Lyapunov}    
 \mathbb{E}\left[ \mathcal{L}(\mathbf{x}^{u}_{i,k+1}) \vert \hat{\mathbf{x}}^{u}_{i,k}, \hat{\mathbf{u}}^{d}_{i,k}, \mathbf{H}^{u}_{i,k},\mathbf{H}^{d}_{i,k}, P^{u}_{i,k}, P^{d}_{i,k} \right] \leq \zeta_{i}  \mathbb{E}\left[ \mathcal{L}(\mathbf{x}^{u}_{i,k}) \vert \hat{\mathbf{x}}^{u}_{i,k} \right] ,
 \end{align} } where the expectation in the right hand side of~\eqref{Future_Lyapunov} is with respect to the plant noise $\mathbf{w}_{k}$ in~\eqref{eq1}, the signal estimation error $\mathbf{v}^{l}_{i,k}$ defined in~\eqref{eq6}, and the signal prediction error $\mathbf{e}^{l}_{i,k}$ defined in~\eqref{eq9+a}. According to the objective function in~\eqref{eq34} and the control stability constraint in~\eqref{Future_Lyapunov}, the control-constrained optimization problem can be formulated as follows: \vspace{-5pt} { \small
  \begin{subequations}\label{eqopt1e}  
  \begin{gather} \label{eqopt1e_1}
  \underset{ \hspace{-310pt}\mathbf{a}_{k}^{l}, \mathbf{P}_{k}^{l}  }{\hspace{-340pt} (\mathcal{P}1) \quad \text{Minimize}} \hspace{-130pt}  \mathcal{C} \left( \{ \bar{\beta}^{l}_{i} \},\{ \bar{\hat{P}}^{l}_{i} \} \right) \\
 \hspace{40pt} \text{subject to:}\quad  0 \leq P^{l}_{i,k} \leq  P^{l}_{max},\qquad \qquad \qquad \qquad \quad \; \; \forall \, l \in \{ u,d \}, i \in \mathcal{M},\, k,    \label{eqopt1e_1a} \\
  \hspace{95pt} \Vert  \mathbf{H}^{l}_{i,k}  \Vert^{2}  \, P^{l}_{i,k} / N_{0} \geq \text{SNR}^{l}_{th}, \qquad \qquad \quad \; \; \; \forall \, l \in \{ u,d \}, i \in \mathcal{M},\, k,   \label{eqopt1e_1b}\\
 \hspace{95pt} \alpha^{l}_{i,k} \in \{ 0,1 \}, \qquad \qquad \qquad \qquad  \qquad \quad \; \forall \, l \in \{ u,d \}, i \in \mathcal{M},\, k,   \label{eqopt1e_1c}\\ 
 \hspace{95pt} \sum^{M}_{i=1} \alpha^{l}_{i,k} \leq 1, \qquad \qquad \qquad \qquad  \qquad \quad \; \forall \, l \in \{ u,d \}, i \in \mathcal{M},\, k,  \label{eqopt1e_1d} \\ 
 \hspace{50pt} \eqref{Future_Lyapunov}, \qquad \qquad \qquad \qquad  \qquad \qquad  \qquad \quad  \forall \;  i \in \mathcal{M},\, k,   \nonumber
 \end{gather}
 \end{subequations} }where $\mathbf{a}_{k}^{l} = \{ \alpha^{l}_{i,k}: \forall l \in \{ u,d \}, \;  i \in \mathcal{M} \}$ and $\mathbf{P}_{k}^{l} = \{ P^{l}_{i,k}:  \forall l \in \{ u,d \}, \;  i \in \mathcal{M} \}$ are the UL-DL scheduling vector at time $k$, and the UL-DL transmission power vector at time $k$, respectively. The constraint in~\eqref{eqopt1e_1a} bounds the UL-DL transmission power allocation of a control system $i$ at time $k$ by the total available transmission power $P^{l}_{max}$, while the constraint in~\eqref{eqopt1e_1b} ensures the communication reliability that is based on SNR or SDR in analog uncoded communications. The constraints in~\eqref{eqopt1e_1c}-\eqref{eqopt1e_1d} ensure at most one transmitter-receiver pair of a control system $i$ is scheduled at time $k$. The constraint in~\eqref{Future_Lyapunov} ensures the state is decreasing along trajectories that satisfies the control stability. It is noted that control stability constraint in~\eqref{Future_Lyapunov} is independent with communication constraints in~\eqref{eqopt1e_1a}-\eqref{eqopt1e_1d}. However, the control stability constraint is affected and determined by the communication and scheduling variables, hence the original control-constrained  problem $\mathcal{P}1$ is rewritten after directly reflecting the communication control relationship of the constraint~\eqref{Future_Lyapunov} in the following lemma.

\begin{lemma}
\label{Stability_Constraints}
 Given  predicted state $\hat{\mathbf{x}}^{u}_{i,k}$,  state prediction error covariance matrix $\mathcal{J}^{u}_{i,k}$, the predicted action $\hat{\mathbf{u}}^{d}_{i,k}$, the action prediction error covariance matrix $\mathcal{J}^{d}_{i,k}$, the channel between the sensor-controller pair $\mathbf{H}^{u}_{i,k}$, the channel between the controller-actuator pair $\mathbf{H}^{d}_{i,k}$, the UL transmission power $P^{u}_{i,k}$, and the DL transmission power $P^{d}_{i,k}$, the control stability constraint in~\eqref{Future_Lyapunov} is satisfied IFF  the following conditions on the transmission indicator variables hold, i.e.,  { \small \begin{align}
 \label{eq32}
 \textstyle{ \underset{K\to \infty}{\lim\sup}  \frac{1}{K} \sum\limits_{k=1}^{K} \xi^{u}_{i,k} \geq  \underset{K\to \infty}{\lim\sup} \frac{1}{K} \sum\limits_{k=1}^{K} \frac{  \Vert \left( \mathbf{A}^{c}_{i} - \zeta_{i} \mathbf{I}_{D}  \right) \hat{\mathbf{x}}^{u}_{i,k} \Vert_{\mathcal{Z}^{\frac{1}{2}}}^{2} +  \text{Tr} \left[ \left( \mathbf{A}^{T}_{i} \mathcal{Z} \mathbf{A}_{i}  - \zeta_{i}  \mathcal{Z} \right) \mathcal{J}^{u}_{i,k} \right]   + \text{Tr} \left[ \mathbf{B}^{T}_{i} \mathcal{Z} \mathbf{B}_{i} \mathcal{J}^{d}_{i,k} \right] + \text{Tr} \left[ \mathcal{Z} \mathbf{W} \right] }{ \text{Tr} \left[  \left( \mathbf{B}_{i} \mathbf{\Phi}_{i} \right)^{T} \mathcal{Z}  \left( \mathbf{B}_{i} \mathbf{\Phi}_{i} \right) \mathcal{J}^{u}_{i,k} \right] -  \text{Tr} \left[  \left( \mathbf{B}_{i} \mathbf{\Phi}_{i} \right)^{T} \mathcal{Z}  \left( \mathbf{B}_{i} \mathbf{\Phi}_{i} \right) \mathbf{V}^{u}_{i,k} \right] }},
 \end{align} }
 \vspace{-10pt} { \small\begin{align} 
 \label{eq33} 
 \textstyle{\underset{K\to \infty}{\lim\sup}  \frac{1}{K} \sum\limits_{k=1}^{K} \xi^{d}_{i,k} \geq  \underset{K\to \infty}{\lim\sup} \frac{1}{K} \sum\limits_{k=1}^{K} \frac{ 
 \Vert \left( \mathbf{A}^{c}_{i} - \zeta_{i} \mathbf{I}_{D}  \right) \hat{\mathbf{x}}^{u}_{i,k} \Vert_{\mathcal{Z}^{\frac{1}{2}}}^{2} +  \text{Tr} \left[ \left( \mathbf{A}^{T}_{i} \mathcal{Z} \mathbf{A}_{i}  - \zeta_{i}  \mathcal{Z} \right) \mathcal{J}^{u}_{i,k} \right]   + \text{Tr} \left[ \mathbf{B}^{T}_{i} \mathcal{Z} \mathbf{B}_{i} \mathcal{J}^{d}_{i,k} \right] + \text{Tr} \left[ \mathcal{Z} \mathbf{W} \right]}{ \text{Tr} \left[ \mathbf{B}_{i}^{T} \mathcal{Z}  \mathbf{B}_{i} \mathcal{J}^{d}_{i,k} \right]  -  \text{Tr}  \left[  \mathbf{B}_{i}^{T} \mathcal{Z}  \mathbf{B}_{i}  \mathbf{V}^{d}_{i,k} \right] }},
\end{align} } 
\vspace{-10pt} { \small\begin{align} 
 \label{eq33e} 
 \textstyle{\underset{K\to \infty}{\lim\sup}  \frac{1}{K} \sum\limits_{k=1}^{K} \xi^{u}_{i,k} \xi^{d}_{i,k} \geq  \underset{K\to \infty}{\lim\sup} \frac{1}{K} \sum\limits_{k=1}^{K} \frac{ 
 \Vert \left( \mathbf{A}^{c}_{i} - \zeta_{i} \mathbf{I}_{D}  \right) \hat{\mathbf{x}}^{u}_{i,k} \Vert_{\mathcal{Z}^{\frac{1}{2}}}^{2} +  \text{Tr} \left[ \left( \mathbf{A}^{T}_{i} \mathcal{Z} \mathbf{A}_{i}  - \zeta_{i}  \mathcal{Z} \right) \mathcal{J}^{u}_{i,k} \right]   + \text{Tr} \left[ \mathbf{B}^{T}_{i} \mathcal{Z} \mathbf{B}_{i} \mathcal{J}^{d}_{i,k} \right] + \text{Tr} \left[ \mathcal{Z} \mathbf{W} \right]}{ 
 \text{Tr} \left[ \left( (\mathbf{B}_{i} \mathbf{\Phi}_{i})^{T} \mathcal{Z}  (\mathbf{B}_{i} \mathbf{\Phi}_{i} ) \right) (\mathcal{J}^{u}_{i,k} - \mathbf{V}^{u}_{i,k})  \right]  + \text{Tr} \left[ \mathbf{B}_{i}^{T} \mathcal{Z}  \mathbf{B}_{i}  (\mathcal{J}^{d}_{i,k} - \mathbf{V}^{d}_{i,k}   \right] }}.
\end{align}} \vspace{-25pt}
 \begin{proof}
 Please refer to Appendix.~\ref{App.lemma2}
 \end{proof}
 \end{lemma} 
 
Note that the conditions on the UL and DL transmission indicator variables in~\eqref{eq32} and~\eqref{eq33}, respectively, ensure the control stability constraint in~\eqref{Future_Lyapunov} in a decoupled scheduling between the UL and DL communications based on the current predicted control and channel states, while the condition on both the UL and DL scheduling variables in~\eqref{eq33e} ensures  control stability in a coupled scheduling between the UL and DL communications. Intuitively, the growth of AoI at a controller/actuator leads to an increasing state/action prediction error increasing due to an outdated training set. Therefore, the transmitter-receiver pair of a control system should be scheduled when it has a reliable communication link and the state/action prediction error is greater than the state/action estimation error to ensure control stability. 

 The actuator is physically decoupled from the centralized scheduler and the controller which is co-located at BS, and the DL indicator variable at the centralized scheduler relies on the action prediction error at the actuator. Hence, the controller leverages another GPR, where the input of this GPR is the discrete-time associated with the generated action by LQR plus the action estimation error as $\bar{\mathbf{u}}^{d_{c}}_{i,k'} = \mathbf{u}^{d}_{i,k'} + \mathbf{v}^{d}_{i,k'} $. As a result of the applied input to this GPR that yields a training set similar to the one at the actuator as $\mathcal{D}^{d_{c}}_{i,n_{d}} = \{( k', \xi^{d}_{i,k'} \,  \bar{\mathbf{u}}^{d_{c}}_{i,k'} )|\, k' = 1, \cdots, n_{d}, i=1,\cdots,M \}$, we obtain the action prediction error similar to the one generated at  actuator side.

 \subsection{ Joint Communication and Control Problem }
 \label{sec3_2.Problem_Reformulation}

 According to the UL and DL transmission indicator variables constraints in~\eqref{eq32}-\eqref{eq33e} that result from the control stability constraint in~\eqref{Future_Lyapunov} in problem $ \mathcal{P}1 $, problem $\mathcal{P}1 $ is rewritten as \vspace{-3pt}  { \small
 \begin{subequations}\label{eqopt2}  
 \begin{gather}
 \label{eqopt2_1}
 \underset{ \hspace{-110pt}\mathbf{a}_{k}^{l}, \mathbf{P}_{k}^{l}  }{\hspace{-140pt} (\mathcal{P}2) \quad \text{Minimize}} \hspace{-30pt}  \mathcal{C} \left( \{ \bar{\beta}^{l}_{i} \}, \{\bar{ \hat{P}}^{l}_{i} \} \right) \\ \hspace{-10pt}
 \text{subject to:}\quad   \bar{\alpha}^{u}_{i}  \geq \bar{\mathcal{G}}_{lb} \left( \mathfrak{m}^{u}_{i,k} \right),\; \; \;  \forall i \in \mathcal{M},   \label{eqopt1_1a}\\
 \hspace{45pt}  \bar{\alpha}^{d}_{i}  \geq \bar{\mathcal{G}}_{lb} \left( \mathfrak{m}^{d}_{i,k} \right),\; \; \; \forall i \in \mathcal{M},   \label{eqopt1_1b}\\
 \hspace{45pt}  \overline{\alpha^{u}_{i} \alpha^{d}_{i}}  \geq \bar{\mathcal{G}}_{lb} \left( \mathfrak{m}_{i,k} \right),\; \forall i \in \mathcal{M},   \label{eqopt1_1c}\\
 \hspace{-15pt}  ~\eqref{eqopt1e_1a}~-~\eqref{eqopt1e_1d},  \nonumber
 \label{eqopt1_1d}
 \end{gather}
 \end{subequations} } where $\bar{\alpha}^{u}_{i}$ and $\bar{\alpha}^{d}_{i}$ are the time-averaged of the UL and DL scheduling variables, respectively, $\mathfrak{m}^{u}_{i,k}$, $\mathfrak{m}^{d}_{i,k}$, $\mathfrak{m}_{i,k}$ are the lower-bound stability of the uplink, downlink, and coupling transmission indicator variables of a control system $i$ at time $k$ in~\eqref{eq32}, \eqref{eq33}, and \eqref{eq33e} respectively. $\bar{\mathcal{G}}_{lb}$ is the time-averaged of the lower-bound function $\mathcal{G}_{lb}$ that is defined as $\mathcal{G}_{lb} \left( .\right) = \max \left[ \min \left( . , 1\right),0 \right] $ to ensure the feasibility of the scheduling constraints. The transmission indicator variable in~\eqref{eq32}-\eqref{eq33e} is only written as a function of the scheduling variables since the SNR indicator function is satisfied in~\eqref{eqopt1e_1b}. The stochastic problem $\mathcal{P}2$ is a mixed-integer non-convex problem where the source of stochasticity is due to the observed channel and predicted state at each time $k$. Moreover, the scheduling decision constraint in~\eqref{eqopt1e_1c} not only depends on its own decision but on all others scheduling decisions. Hence, a dynamic control algorithm is proposed in the next section  to find the optimal scheduling vector and the optimal transmission power vector of problem $\mathcal{P}2$ utilizing Lyapunov optimization framework.

 \section{Dynamic Control Algorithm Using Lyapunov Optimization}
  \label{sec4.Lyapunov_Optimization}
 In this section, we propose a dynamic control algorithm using the stochastic Lyapunov optimization framework to solve problem $\mathcal{P}2$. However, the problem involves minimizing a weighted sum of non-decreasing concave functions of the time-averaged AoI and transmission power. Based on the dynamic stochastic optimization theory~\cite{neely2010stability}, it can be transformed into an equivalent problem that involves minimizing a time-averaged cost function of instantaneous AoI and transmission power. This transformation is achieved through the use of non-negative auxiliary variables $\gamma^{\beta^{l}}_{i,k}$ and $\gamma^{P^{l}}_{i,k}$ and corresponding virtual queues $Q^{\beta^{l}}_{i,k}$ and $Q^{P^{l}}_{i,k}$ with queue dynamics as \vspace{-2pt}   { \small \begin{subequations} 
\begin{gather}
\label{eq37}
Q^{\beta^{l}}_{i,k+1} = \max \Big\{Q^{\beta^{l}}_{i,k} - \gamma^{\beta^{l}}_{i,k}, 0  \Big\} + \beta^{l}_{i,k},  \qquad \forall l \in \{ u, d \}, i \in \mathcal{M}, k, \\
\label{eq39}
Q^{P^{l}}_{i,k+1} = \max \Big\{Q^{P^{l}}_{i,k} - \gamma^{P^{l}}_{i,k}, 0  \Big\} + \hat{P}^{l}_{i,k} , \qquad \forall l \in \{ u, d \}, i \in \mathcal{M}, k, 
\end{gather}
\end{subequations}} where $\hat{P}^{l}_{i,k}$  will be optimized at each time $k$. Then, the transformed problem is given as  { \small
\begin{subequations}\label{eqopt2} 
 \begin{gather} \label{eqopt2_1}
 \underset{ \hspace{-320pt}\mathbf{a}_{k}^{l}, \mathbf{P}_{k}^{l}, \mathbf{r}^{\beta^{l}}, \mathbf{r}^{P^{l}}  }{\hspace{-360pt} (\mathcal{P}3) \quad \text{Minimize}} \hspace{-120pt}  \overline{\mathcal{C} \left( \{ \gamma^{\beta^{l}}_{i,k} \} , \{ \gamma^{P^{l}}_{i,k} \} \right)} \\
 \hspace{-90pt}  \text{subject to:}\qquad   \bar{\beta}^{l}_{i} \leq  \bar{\gamma}_{i}^{\beta^{l}},\qquad \qquad \quad \forall l \in \{ u, d \}, i \in \mathcal{M}   \label{eqopt2_1a} \\
 \hspace{-25pt}  \bar{\hat{P}}^{l}_{i} \leq  \bar{\gamma}_{i}^{P^{l}},\qquad  \qquad \quad \forall l \in \{ u, d \}, i \in \mathcal{M}  \label{eqopt2_1b} \\
\hspace{-15pt}   1  \leq \gamma_{i,k}^{\beta^{l}} \leq B_{max}, \qquad \forall l \in \{ u, d \}, i \in \mathcal{M}, k  \label{eqopt2_1c}\\
\hspace{-15pt}   0  \leq \gamma_{i,k}^{P^{f}} \leq P^{l}_{max}, \qquad \forall l \in \{ u, d \}, i \in \mathcal{M}, k   \label{eqopt2_1d}\\
\hspace{-80pt} ~\eqref{eqopt1e_1a}-\eqref{eqopt1e_1d},~\eqref{eqopt1_1a}-\eqref{eqopt1_1c},  \nonumber \label{eqopt2_1f}
\end{gather} 
\end{subequations} } where $\mathbf{r}_{k}^{\beta^{l}} = \{  \gamma^{\beta^{l}}_{i,k} : l \in \{u,d \},  i \in \mathcal{M} \}$ and $\mathbf{r}_{k}^{P^{l}} = \{  \gamma^{P^{l}}_{i,k} : l \in \{u,d \}, i \in \mathcal{M} \}$  are the vectors of the introduced auxiliary variables. The constraints in~\eqref{eqopt2_1c} and~\eqref{eqopt2_1d}  are introduced to bound the auxiliary variables. These constraints can be satisfied by ensuring the stability of their virtual queues, since the lower-bound of these constraints can be viewed as the arrival rate of their virtual queues, while the upper-bound can be viewed as the service rate of such virtual queues~\cite{neely2010stability}. Following~\cite{neely2010stability}, the problem $\mathcal{P}2$ and the transformed problem $ \mathcal{P}3$ are equivalent in which the optimal solution of $\mathcal{P}3$ can be directly turned into an optimal solution of $\mathcal{P}2$. 
 
 To handle  UL and DL scheduling variables constraints in~\eqref{eqopt1_1a}$-$\eqref{eqopt1_1c} associated with the control stability constraint in~\eqref{Future_Lyapunov}, the virtual queues $Q^{C^{l}}_{i,k}$ and $Q^{C}_{i,k}$ are introduced for all control systems whose dynamics are {\small \begin{subequations}  
 \begin{gather}
\label{eq43}
 Q^{C^{l}}_{i,k+1} = \max \{Q^{C^{l}}_{i,k} - \alpha^{l}_{i,k}, 0  \} + \mathcal{G}_{lb} ( m^{l}_{i,k} ),\qquad \forall l \in \{ u, d \}, i \in \mathcal{M}, k,  \\
\label{eq43e}
 Q^{C}_{i,k+1} = \max \{Q^{C}_{i,k} - \alpha^{u}_{i,k} \alpha^{d}_{i,k} , 0  \} + \mathcal{G}_{lb} ( m_{i,k} ),\qquad \qquad \; \; \; \forall i \in \mathcal{M}, k, \end{gather}  \end{subequations} } where $\alpha^{l}_{i,k}$ will be optimized at each time $k$. The constraints in~\eqref{eqopt1_1a}$-$\eqref{eqopt1_1b} can be satisfied, if their virtual queues are mean-rate stable, i.e., their time-averaged arrival rate is not larger than its time-averaged service rate~\cite{neely2010stability}. At this point, the dynamic stochastic optimization is applied to solve the transformed problem  $ \mathcal{P}3 $, which minimizes a weighted sum of the time-averaged cost function of instantaneous AoI and transmission power subject to the virtual queues stability constraints and the original problem constraints in~\eqref{eqopt1e_1a}$-$\eqref{eqopt1e_1d}. In this regard, we define $\mathbf{Q}^{\beta^{l}}_{k}$, $\mathbf{Q}^{P^{l}}_{k}$, $\mathbf{Q}^{C}_{k}$, and $\mathbf{Q}^{C^{l}}_{k}$ as a vector of all virtual queues $Q^{\beta^{l}}_{i,k}$, $ Q^{P^{l}}_{i,k}$, $ Q^{C}_{i,k}$, and $ Q^{C^{l}}_{i,k}$ for all control systems, respectively. We denote the combined queue vector of all virtual queues at time $k$ by ${\scriptstyle \mathcal{X}_{k} = \left[ \mathbf{Q}^{\beta^{l}}_{k}, \mathbf{Q}^{P^{l}}_{k}, \mathbf{Q}^{C}_{k},  \mathbf{Q}^{C^{l}}_{k} \right]}$, and express the conditional Lyapunov drift-plus-penalty as \begin{equation}  \small
\label{eq46}
\Delta \left(  \mathcal{X}_{k}  \right) = \mathbb{E} \left[ \mathcal{L}\left( \mathcal{X}_{k+1} \right)  - \mathcal{L}\left( \mathcal{X}_{k} \right) + V \mathcal{C} \left( \{  \gamma^{\beta^{l}}_{i,k} \}, \{\gamma^{P^{l}}_{i,k} \} \right) \Big| \mathcal{X}_{k} \right],
\end{equation}  where $\mathcal{L}\left( \mathcal{X}_{k} \right) $ is the quadratic Lyapunov function of $\mathcal{X}_{k}$ that measures the virtual queues congestion in a scalar metric and is defined as $\mathcal{L}\left( \mathcal{X}_{k} \right) =  \frac{1}{2}  \sum^{M}_{i=1} \Big[  (Q^{\beta^{l}}_{i,k})^{2} + (Q^{P^{l}}_{i,k})^{2} +  (Q^{C}_{i,k} )^{2} +  (Q^{C^{l}}_{i,k} )^{2}  \Big]$.  $V \geq 0 $ controls the trade-off between minimizing the objective function and stabilizing the virtual queues. Subsequently, plugging the inequalities $\left( \max \left[  a - b, 0 \right] + c \right)^{2} \leq a^{2} + b^{2} + c^{2} - 2 a\left( b -c \right)$, $\forall a,b,c \geq 0 $, $\left( \max \left( a,0 \right) \right)^{2}  \leq  a^{2}$, and all  virtual queue dynamics into~\eqref{eq46}, we  derive  {\small\begin{equation} 
\label{eq47}
\begin{aligned}
\eqref{eq46} & \leq  {B} + \mathbb{E} \Big[ \sum^{M}_{i=1} \left(  V \omega_{\beta_{l}} \mathcal{G}_{\beta} \left( \gamma^{\beta^{l}}_{i,k} \right) - Q^{\beta^{l}}_{i,k}  \gamma^{\beta^{l}}_{i,k}  \right) \vert \mathcal{X}_{k}  \Big] + \mathbb{E} \Big[ \sum^{M}_{i=1} \left(  V \omega_{P_{l}} \mathcal{G}_{P} \left( \gamma^{P^{l}}_{i,k} \right) - Q^{P^{l}}_{i,k}  \gamma^{P^{l}}_{i,k}  \right)  \vert \mathcal{X}_{k} \Big]  \\ & \qquad \, + \mathbb{E} \Big[ \sum^{M}_{i=1}  Q^{\beta^{l}}_{i,k} \beta^{l}_{i,k}  \vert \mathcal{X}_{k}   \Big]  + \mathbb{E} \Big[ \sum^{M}_{i=1}  Q^{P^{l}}_{i,k} \hat{P}^{l}_{i,k}  \vert \mathcal{X}_{k} \Big] - \mathbb{E} \Big[ \sum^{M}_{i=1}  Q^{C^{l}}_{i,k} \left( \alpha^{l}_{i,k}   - \mathcal{G}_{lb} ( m^{l}_{i,k} ) \right)  \vert \mathcal{X}_{k} \Big]  \\ & \qquad \, - \mathbb{E} \Big[ \sum^{M}_{i=1}  Q^{C}_{i,k} \left( \alpha^{u}_{i,k} \alpha^{d}_{i,k}   - \mathcal{G}_{lb} ( m_{i,k} ) \right)  \vert \mathcal{X}_{k} \Big].
\end{aligned}
\end{equation} } The constant B details in~\eqref{eq47} is omitted since it does not affect the system performance in the Lyapunov optimization. A solution to $\mathcal{P}3$ can be obtained by minimizing the upper-bound~\eqref{eq47} at each time as 
\begin{subequations} \small \label{eqopt3_1}
\begin{gather}
 \underset{ \hspace{25pt}\mathbf{a}^{l}, \mathbf{P}^{l}, \mathbf{r}^{\beta^{l}}, \mathbf{r}^{P^{l}} }{\hspace{-10pt} (\mathcal{P}4)  \quad  \text{Minimize}}  \quad   \sum^{M}_{i=1} \Bigg[ \left(  V \omega_{\beta_{l}} \mathcal{G}_{\beta} \left( \gamma^{\beta^{l}}_{i,k} \right) - Q^{\beta^{l}}_{i,k}  \gamma^{\beta^{l}}_{i,k}  \right)  +   \left(  V \omega_{P_{l}} \mathcal{G}_{P} \left( \gamma^{P^{l}}_{i,k} \right) - Q^{P^{l}}_{i,k}  \gamma^{P^{l}}_{i,k}  \right) +   Q^{\beta^{l}}_{i,k} \beta^{l}_{i,k}  \\  \qquad \qquad  \qquad \qquad \qquad   + Q^{P^{l}}_{i,k} \hat{P}^{l}_{i,k}   -   Q^{C^{l}}_{i,k} \left( \alpha^{l}_{i,k}  -  \mathcal{G}_{lb} (m^{l}_{i,k})  \right)  -   Q^{C}_{i,k} \left( \alpha^{u}_{i,k} \alpha^{d}_{i,k}  -  \mathcal{G}_{lb} (m_{i,k})  \right) \Bigg] \nonumber \\  \hspace{-160pt}  \text{subject to:} \qquad ~\eqref{eqopt1e_1a}-\eqref{eqopt1e_1d}~\text{and}~\eqref{eqopt2_1c}-\eqref{eqopt2_1d}. \nonumber
 \end{gather}
\end{subequations} The optimality of problem $\mathcal{P}4$ is asymptotically approached by increasing $V$~\cite{neely2010stability}. Since the problem $\mathcal{P}4$ is of separable structure, which motivate us to determine the AoI auxiliary vector $\mathbf{r}^{\beta^{l}}$, transmission power auxiliary vector $\mathbf{r}^{P^{l}}$, scheduling vector $\mathbf{a}^{l}$, and transmission power vector $\mathbf{P}^{l}$ in an alternative optimization form. Hence, the overall minimization problem $\mathcal{P}4$ can be decomposed into two separate sub-problems that can be solved concurrently with the observation of the virtual queues, control, and channel states. 

\subsubsection{Auxiliary Variable Sub-Problems} \label{sec4_1_1} The first decomposed sub-problem is the AoI auxiliary sub-problem, while the second decomposed sub-problem is the transmission power sub-problem. Since the auxiliary variables of such problems are separated and independent among different control systems, their minimization sub-problems can be decoupled to be computed for each control system separately as the following convex problems \vspace{-2pt}
 { \small \begin{subequations}\label{eqopt4} 
\begin{gather} \label{eqopt4_1}
 \underset{\hspace{45pt}  \gamma^{\beta^{l}}_{i,k}}{ (\mathcal{P}4.1) \quad \text{Minimize}} \quad   V \omega_{\beta_{l}} \mathcal{G}_{\beta} \left( \gamma^{\beta^{l}}_{i,k} \right) - Q^{\beta^{l}}_{i,k}  \gamma^{\beta^{l}}_{i,k}    \\
\hspace{20pt} \text{subject to:}\quad  \quad   1  \leq \gamma_{i,k}^{\beta^{l}} \leq B_{max}, \label{eqopt4_1a} 
\end{gather} \vspace{-8pt}
\end{subequations} } \vspace{-3pt} { \small \begin{subequations}\label{eqopt6} 
\begin{gather} \label{eqopt6_1}
 \underset{\hspace{45pt}  \gamma^{P^{l}}_{i,k}}{ (\mathcal{P}4.2) \quad \text{Minimize}} \quad   V \omega_{P_{l}} \mathcal{G}_{P} \left( \gamma^{P^{l}}_{i,k} \right) - Q^{P^{l}}_{i,k}  \gamma^{P^{l}}_{i,k}  \\ 
\hspace{20pt} \text{subject to:}\quad  \quad   0  \leq \gamma_{i,k}^{P^{l}} \leq P^{l}_{max}. \label{eqopt6_1a} 
\end{gather}
\end{subequations} }  The optimal AoI auxiliary variables are obtained by differentiating the objective functions of these problems. Let $\mathcal{A} ( \gamma^{\beta^{l}} ) = V \omega_{\beta_{l}} \log  ( 1 + \gamma^{\beta^{l}} )  - Q^{\beta^{l}}_{i,k}  \gamma^{\beta^{l}}_{i,k}$ and $\gamma^{\beta^{l^{*}}}_{i,k}$ denotes the solution of $\mathcal{A} ( \gamma^{\beta^{l}} )$ as $\acute{\mathcal{A}} ( \gamma^{\beta^{l}} ) = \frac{V \omega_{\beta_{l}}}{\left( 1 + \gamma^{\beta^{l}}_{i,k} \right)} - Q^{\beta^{l}}_{i,k}  = 0$, the optimal AoI auxiliary variable of~$\mathcal{P}4.1$ is given as \begin{equation} \small
 \label{eq52}
\gamma^{\beta^{l^{*}}}_{i,k} = \min \left\lbrace \max \left\lbrace  \frac{ V \omega_{\beta_{l}} - Q^{\beta^{l}}_{i,k} }{ Q^{\beta^{l}}_{i,k}},1 \right\rbrace, B_{max} \right\rbrace,\qquad \forall l \in \{ u,d \}, i \in \mathcal{M}, k.
\end{equation}  Similarly, by letting $\mathcal{A} ( \gamma^{P^{l}} ) = V \omega_{P^{l}} \log  ( 1 + \gamma^{P^{l}} )  \,  -  \, Q^{P^{l}}_{i,k}  \gamma^{P^{l}}_{i,k}$ and $\gamma^{P^{l^{*}}}_{i,k}$ denotes the solution of $\mathcal{A} ( \gamma^{P^{l}} )$ as $\acute{\mathcal{A}} ( \gamma^{P^{l}} ) = \frac{V \omega_{P^{l}}}{\left( 1 + \gamma^{P^{l}}_{i,k} \right)} - Q^{P^{l}}_{i,k}  = 0$, the optimal transmission power auxiliary variable of~$\mathcal{P}4.2$ is given as  \begin{equation} \small
\label{eq54}
\gamma^{P^{l^{*}}}_{i,k} = \min \left\lbrace \max \left\lbrace \frac{ V \omega_{P^{l}} - Q^{P^{l}}_{i,k} }{ Q^{P^{l}}_{i,k}},0 \right\rbrace, P^{l}_{max} \right\rbrace, \qquad \forall l \in \{ u,d \}, i \in \mathcal{M}, k.
\end{equation}  \subsubsection{Scheduling Decision and Transmission Power Sub-Problems} \label{sec4_1_2} The optimal  UL-DL scheduling variables and the optimal UL-DL transmission power variables are obtained by minimizing the remaining terms of the objective function of problem $\mathcal{P}4$ at each time subject to the scheduling and transmission power constraints in~\eqref{eqopt1e_1a}$-$\eqref{eqopt1e_1d}, which is expressed as  
{\small \begin{subequations}\label{eqopt8} 
\begin{gather} \label{eqopt8_1} \small
\underset{\hspace{40pt} \mathbf{a}^{l}_{i,k}, \mathbf{P}^{l}_{i,k}}{ (\mathcal{P}4.3) \quad  \text{Minimize}} \; \sum^{M}_{i=1} \left[  Q^{\beta^{l}}_{i,k} \beta^{l}_{i,k} + Q^{P^{l}}_{i,k} \, \hat{P}^{l}_{i,k} -   Q^{C^{l}}_{i,k} \left( \alpha^{l}_{i,k} - \mathcal{G}_{lb} (m^{l}_{i,k})  \right) - Q^{C}_{i,k} \left( \alpha^{u}_{i,k} \alpha^{d}_{i,k} - \mathcal{G}_{lb} (m_{i,k})  \right) \right] \\
\hspace{-250pt} \text{subject to:}\quad  ~\eqref{eqopt1e_1a}-\eqref{eqopt1e_1d},  \nonumber \label{eqopt8_1a}
\end{gather}
\end{subequations} } which is a mixed-integer non-convex problem. Due to the  complexity of   exhaustive search for finding the optimal solution, we propose a low-complexity two-stage sequential optimization strategy to find a sub-optimal solution to the joint power allocation and scheduling assignment problem. This strategy firstly obtains the UL and DL transmission power variables,  followed by the UL and DL scheduling variables. The optimal UL and DL transmission power for each control system, determined by solving the following power allocation problem  { \small \begin{subequations}\label{eqopt11} 
\begin{gather} \label{eqopt11_1}  
\underset{\hspace{15pt} \mathbf{P}^{u}_{k},  \mathbf{P}^{d}_{k}}{ \hspace{-20pt} (\mathcal{P}4.4) \quad \text{Minimize}} \; \sum^{M}_{i=1}  \alpha^{u}_{i,k} \left[ \mathcal{Q}^{S_{1}}_{i,k}  + Q^{P^{u}}_{i,k} P^{u}_{i,k} \right] +  \alpha^{d}_{i,k} \left[ \mathcal{Q}^{C_{1}}_{i,k}  + Q^{P^{d}}_{i,k} P^{d}_{i,k} \right] - \alpha^{u}_{i,k} \alpha^{d}_{i,k} Q^{C}_{i,k} + \mathcal{Q}^{l_{1}}_{i,k},\\
\hspace{-100pt} \text{subject to:}\quad   
   \frac{ \text{SNR}^{u}_{th} \, {N_{0}} }{ \Vert \mathbf{H}^{u}_{i,k} \Vert^{2} }  \leq  P^{u}_{i,k} \leq  P^{u}_{max}, \qquad  \forall i \in \mathcal{M},k \label{eqopt11_1a}
 \\ \hspace{-40pt}  \frac{ \text{SNR}^{d}_{th} \, {N_{0}} }{ \Vert \mathbf{H}^{d}_{i,k} \Vert^{2} }  \leq  P^{d}_{i,k} \leq  P^{d}_{max}, \qquad  \forall i \in \mathcal{M},k,  \label{eqopt11_1b}
\end{gather}
\end{subequations} } where $\mathcal{Q}^{S_{1}}_{i,k} = -Q^{\beta^{u}}_{i,k} \beta^{u}_{i,k-1}  - Q^{C^{u}}_{i,k}$, $\mathcal{Q}^{C_{1}}_{i,k} = -Q^{\beta^{d}}_{i,k} \beta^{d}_{i,k-1}  - Q^{C^{d}}_{i,k}$, and $\mathcal{Q}^{l_{1}}_{i,k} = Q^{C^{u}}_{i,k} \mathcal{G}_{lb}(m^{u}_{i,k}) + Q^{C^{d}}_{i,k} \mathcal{G}_{lb}(m^{d}_{i,k}) + Q^{C}_{i,k} \mathcal{G}_{lb}(m_{i,k}) + Q_{i,k}^{\beta^{u}} \left( 1 + \beta^{u}_{i,k-1} \right)  + Q_{i,k}^{\beta^{d}} \left( 1 + \beta^{d}_{i,k-1} \right) $  are the constant terms defined in the objective function of~$\mathcal{P}4.4$. The above problem~$\mathcal{P}4.4$ is a generalized min-weight problem, which can be decoupled into a series of independent sub-problems for each control system separately. Hence, the optimal UL-DL transmission power variables are given as \begin{equation} \small 
\label{eq58}
P^{l^{*}}_{i,k}  =  \left\{ \begin{array}{cc} \frac{\text{SNR}^{l}_{th} N_{0}}{\Vert \mathbf{H}^{l}_{i,k} \Vert^{2}}, &\text{if} \; Q^{P^{l}}_{i,k} \geq 0 \\  P^{l}_{max}, & \mbox{if} \; Q^{P^{l}}_{i,k} < 0.   
\end{array}
\right.
\end{equation}  

Given the optimal UL and DL transmission power variables in~\eqref{eq58}, the optimal UL and DL scheduling variables for each control system that has a control state/action to transmit are obtained by solving the following scheduling assignment problem { \small  \begin{subequations}\label{eqopt10} 
\begin{gather} \label{eqopt10_1} 
\underset{\hspace{40pt} \mathbf{a}^{u}_{k},  \mathbf{a}^{d}_{k}}{(\mathcal{P}4.5) \quad \text{Minimize}} \; \sum^{M}_{i=1}  \alpha^{u}_{i,k} \left[ \mathcal{Q}^{S_{1}}_{i,k} + Q^{P^{u}}_{i,k} P^{u^{*}}_{i,k} \right] +  \alpha^{d}_{i,k} \left[ \mathcal{Q}^{C_{1}}_{i,k}  + Q^{P^{d}}_{i,k} P^{d^{*}}_{i,k} \right] - \alpha^{u}_{i,k} \alpha^{d}_{i,k} Q^{C}_{i,k} + \mathcal{Q}^{l_{1}}_{i,k}, \\
\hspace{-220pt} \text{subject to:}\quad   
~\eqref{eqopt1e_1c}-\eqref{eqopt1e_1d}.  \nonumber   \label{eqopt10_1a}  
\end{gather}
\end{subequations} } The optimal UL and DL scheduling variables are obtained as follows:  \begin{equation}  \small
\label{eq_sol1}
 \alpha^{u^{*}}_{i,k} \, \& \, \alpha^{d^{*}}_{i,k} =  \left\{ \begin{array}{cc} \alpha^{u}_{j_1,k} =1, \; \alpha^{d}_{j_2,k} =1, &\text{if} \; \mathbb{Q}^{1}_{j_1,k} + \mathbb{Q}^{2}_{j_2,k} < \mathbb{Q}^{1}_{j_3,k} + \mathbb{Q}^{2}_{j3,k} + \mathbb{Q}^{3}_{j_3,k} \\  \alpha^{u}_{j_3,k} =1, \; \alpha^{d}_{j_3,k} =1, &\text{if} \; \mathbb{Q}^{1}_{j_1,k} + \mathbb{Q}^{2}_{j_2,k} > \mathbb{Q}^{1}_{j_3,k} + \mathbb{Q}^{2}_{j_3,k} + \mathbb{Q}^{3}_{j_3,k} \\ \alpha^{u}_{j,k} =0, \; \alpha^{d}_{j,k} =0, &  \forall j \notin \{j_1 \, \& \, j_2 || j_3 \,\},
\end{array}
\right.    
\end{equation}  where $\mathbb{Q}_{i,k}^{1} = \mathcal{Q}^{S_{1}}_{i,k} + Q^{P^{u}}_{i,k} P^{u^{*}}_{i,k}$, $\mathbb{Q}_{i,k}^{2} = \mathcal{Q}^{C_{1}}_{i,k} + Q^{P^{d}}_{i,k} P^{d^{*}}_{i,k}$, and $\mathbb{Q}_{i,k}^{3} = - Q^{C}_{i,k}$ are the terms defined in the objective function of problem $\mathcal{P}4.5$. Moreover,  $j_1 = \argminA_{i \in \mathcal{M}}  \mathbb{Q}^{1}_{i,k}$, $j_2 = \argminA_{i \in \mathcal{M}}  \mathbb{Q}^{2}_{i,k}$, and $j_3 = \argminA_{i \in \mathcal{M}} ( \mathbb{Q}^{1}_{i,k} + \mathbb{Q}^{2}_{i,k} + \mathbb{Q}^{3}_{i,k}  )$ are control system indices. 



\section{Simulation Results and Discussions}
\label{Sec5.Simulation_Results}

In this section, the performance of the proposed stability-aware scheduling algorithm is investigated in an inverted-pendulum on a cart system with  $M=2$, and $M = 20$ inverted-pendulums, respectively. Each inverted-pendulum system is described by a four-dimensional state vector as $\mathbf{x}^{u}_{i,k} = \left[ x_{i,k}, \dot{x}_{i,k}, \theta_{i,k}, \dot{\theta}_{i,k}  \right]$, where $x_{i,k}$ represents the cart's position along the horizontal axis, $\dot{x}_{i,k}$ represents the cart's velocity, $ \theta_{i,k}$ represents the pendulum angle along the vertical axis, and $\dot{\theta}_{i,k} $ represents the pendulum's angular velocity. The initial state of the control systems $i$ is $\mathbf{x}^{u}_{i,0} = \left[ 0 \quad 0 \quad  0.1\quad 0 \right]^{T}$. The action $\mathbf{u}^{a}_{i,k}$ is the horizontal force applied on the linear cart. By applying a zeroth-order with a state sampling of $10$ms on the continuous dynamics of the inverted-pendulum system and linearizing around the pendulum up-position, i.e., $\theta_{i,k} = 0$, we obtain the following discrete-time linear dynamics matrices~\cite{eisen2019control},
{ \small \begin{align}
\label{eq66}
\mathbf{A}_{i} = \left[ \begin{array}{cccc}  
1 & 0         &  0     & 0    \\
0 & 2.055     & -0.722 & 4.828 \\
0 & 0.023     & 0.91 & 0.037 \\
0 & 0.677     & -0.453 & 2.055 \\
\end{array} \right], \mathbf{B}_{i} = \left[ \begin{array}{c}  
0.034  \\
0.168  \\
0.019  \\
0.105  \\
\end{array} \right], 
\end{align} } 

Since $\mathbf{A}_{i}$'s largest eigenvalues $\left\lbrace 3.85, 0.42, 0.92, 1.00 \right\rbrace$ is greater than unity, the inverted-pendulum is unstable without an appropriate control action~\cite{oppenheim2010introduction}. To stabilize the control system, the feedback gain matrix $\mathbf{\Phi}_{i}$ is calculated at the controller based on LQR in~\eqref{eq11}. The rest of the simulation parameters are  $P^{l}_{max} = 20 \, \text{dBm}$,  $N_{0} = -20 \, \text{dBm}$, $ \zeta_{i} = 0.01$,  $V =1000$,  $\omega_{\beta} = 1$, $\omega_{P} = 1$, $ h_{k} = 1$, $ h_{q} = 1$, $ \mu  = 1$, and $\sigma_{n}^{2} = 0.01$.

The performance of the proposed stability-aware scheduling method is compared versus five scheduling baselines. In \textit{Baseline}~$1$. (\textbf{Round-Robin Scheduling}), each sensor/controller periodically transmits its state/action over a wireless channel with fixed transmission power and a predefined repeating order~\cite{hespanha2007survey,schenato2007foundations}. In \textit{Baseline}~$2$. (\textbf{Opportunistic Scheduling}), the sensor/controller is scheduled under favorable channel conditions. Otherwise, the controller/actuator applies the last received state/action~\cite{xu2013stability,liu2003framework}. In \textit{Baseline}~$3$, (\textbf{Event-triggered Scheduling without FDMA}), one control system  with the largest state discrepancy, i.e., the difference between the current predicted state using Kalman filtering and the previous received/predicted state is larger than a predefined threshold, is scheduled at each time to transmit the state/action with fixed transmission power~\cite{cervin2008scheduling,postoyan2011event}. In \textit{Baseline}~$4$. (\textbf{Event-triggered Scheduling with FDMA}), each sensor/controller transmits its state/action with fixed transmission power based on its stability condition, i.e.,  difference between the current and previous states is less than a predefined threshold, using FDMA~\cite{branicky2002scheduling}. In \textit{Baseline}~$5$. (\textbf{Ideal Control Scheduling}), all control systems simultaneously transmit their states/actions with ultra-low latency and  high reliability over perfect channels~\cite{schenato2007foundations}. Results are collected over ten independent simulation runs, and each simulation is run for  $K = 90$ discrete-time steps.

\begin{figure}[htb!]
  \centering
\subfigure[Average Pendulum Angle with $M=2$.\label{fig20.a}]{\includegraphics[width=.4\linewidth]{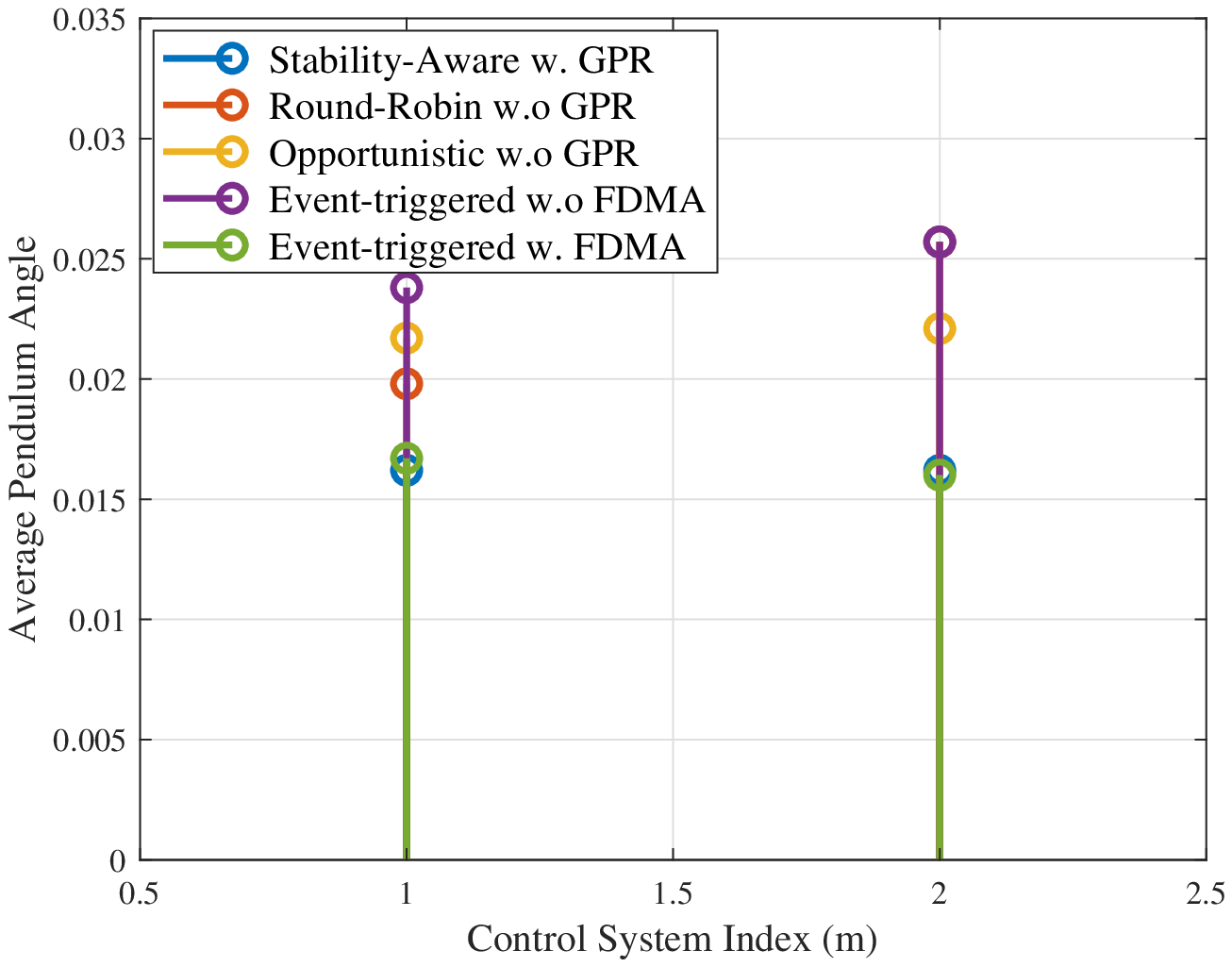}} 
\subfigure[Average Pendulum Angle with $M=20$.\label{fig20.b}]{\includegraphics[width=.4\linewidth]{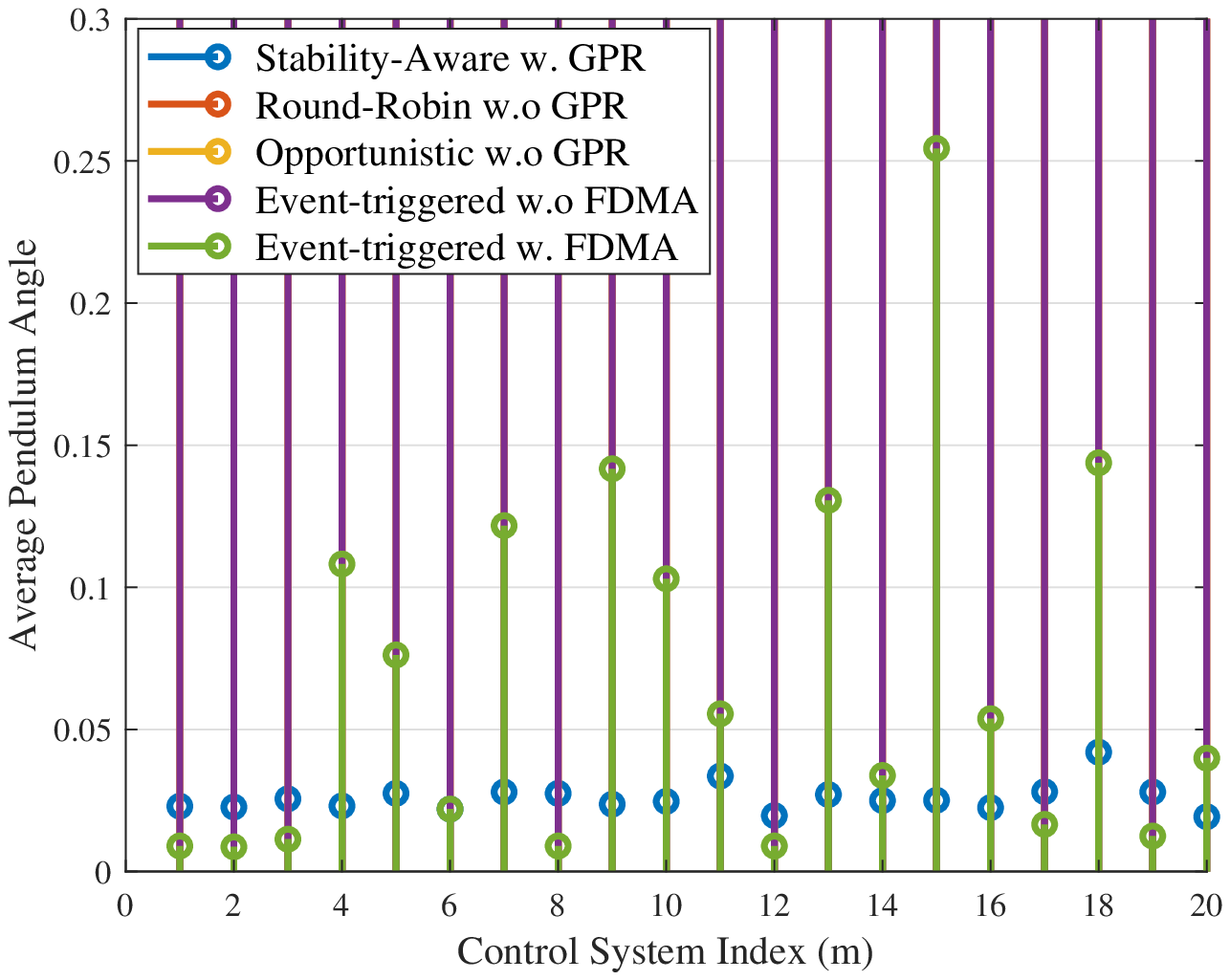}} 
  \caption{Average pendulum angle to the vertical center, i.e., control error with $M=2$ and $M=20$ pendulum angle systems using the proposed stability-aware, round-robin, opportunistic, event-triggered with and without FDMA. } \label{fig20}
\end{figure}


\noindent\textbf{Average Control Error Vs. Control Systems}.
\quad Fig.~\ref{fig20} illustrates the average pendulum angle to the vertical center of each control system, i.e., the average control error of each control system, during $90$ control time steps. As shown in Fig.\ref{fig20.a}, the proposed and baseline scheduling methods, assuming a low number of control systems $\left( M=2 \right)$, can keep all the pendulums upright. Moreover, the proposed stability-Aware with GPR and the event-triggered with FDMA  keep both pendulums close to zero  ensuring  both pendulums have the same control performance. This is because the proposed solution adapts to both channel and control states, i.e., the control system is scheduled if it has a favorable channel condition and unstable control state needs to be stabilized. Meanwhile, the event-triggered with FDMA scheduling has approximately the same performance at the cost of wasting wireless communication resources by transmitting with a fixed transmission power and a high communication rate. The opportunistic scheduler without GPR  has better control performance compared to the event-triggered without FDMA and the round-robin methods leveraging channel state in scheduling compared to the scheduling baselines. This in turn reflects the connection between the state estimation stability and control stability.  

Fig.~\ref{fig20.b} plots the average control error of each control system for the proposed and baseline scheduling methods.  In large numbers of control systems the proposed stability-aware with GPR and the event-triggered with FDMA scheduling methods keep all pendulum upright, unlike the  baselines. Unlike the scheduling baselines except the event-triggered with FDMA scheduling wherein at most one control system is scheduled each time due to the limited bandwidth, our proposed scheduling allows all control systems to operate simultaneously even without receiving either the current state or action, highlighting the effectiveness of GPRs at the controller and actuator  thereby improving communication efficiency and control stability. Moreover, it maintains the GPR prediction credibility, achieving control stability. Meanwhile, the event-triggered with FDMA scheduling keeps some pendulums upright at the cost of frequent transmissions by equally dividing the available bandwidth between the control systems, such that each control system receives a fixed fraction $f_{i}$ of the total capacity $f_{i} = \text{BW}/ M$, affecting transmission latency. 

\begin{figure}[htb!]
  \centering
  \subfigure[Communication rate of sensing link.\label{Fig21.a}]{\includegraphics[width=.4\linewidth]{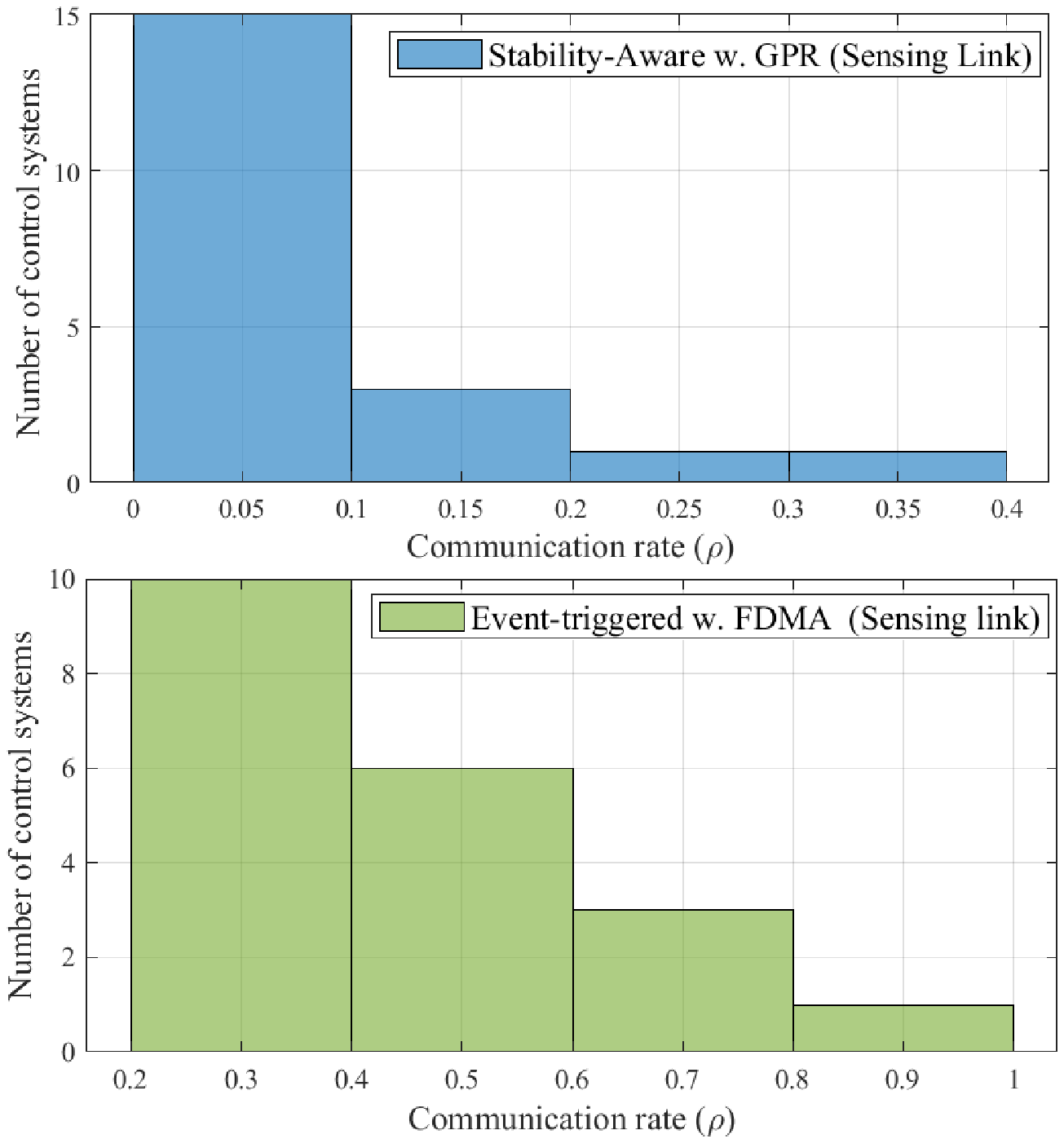} }
  \subfigure[Communication rate of actuating link.  \label{Fig21.b}]{\includegraphics[width=.4\linewidth]{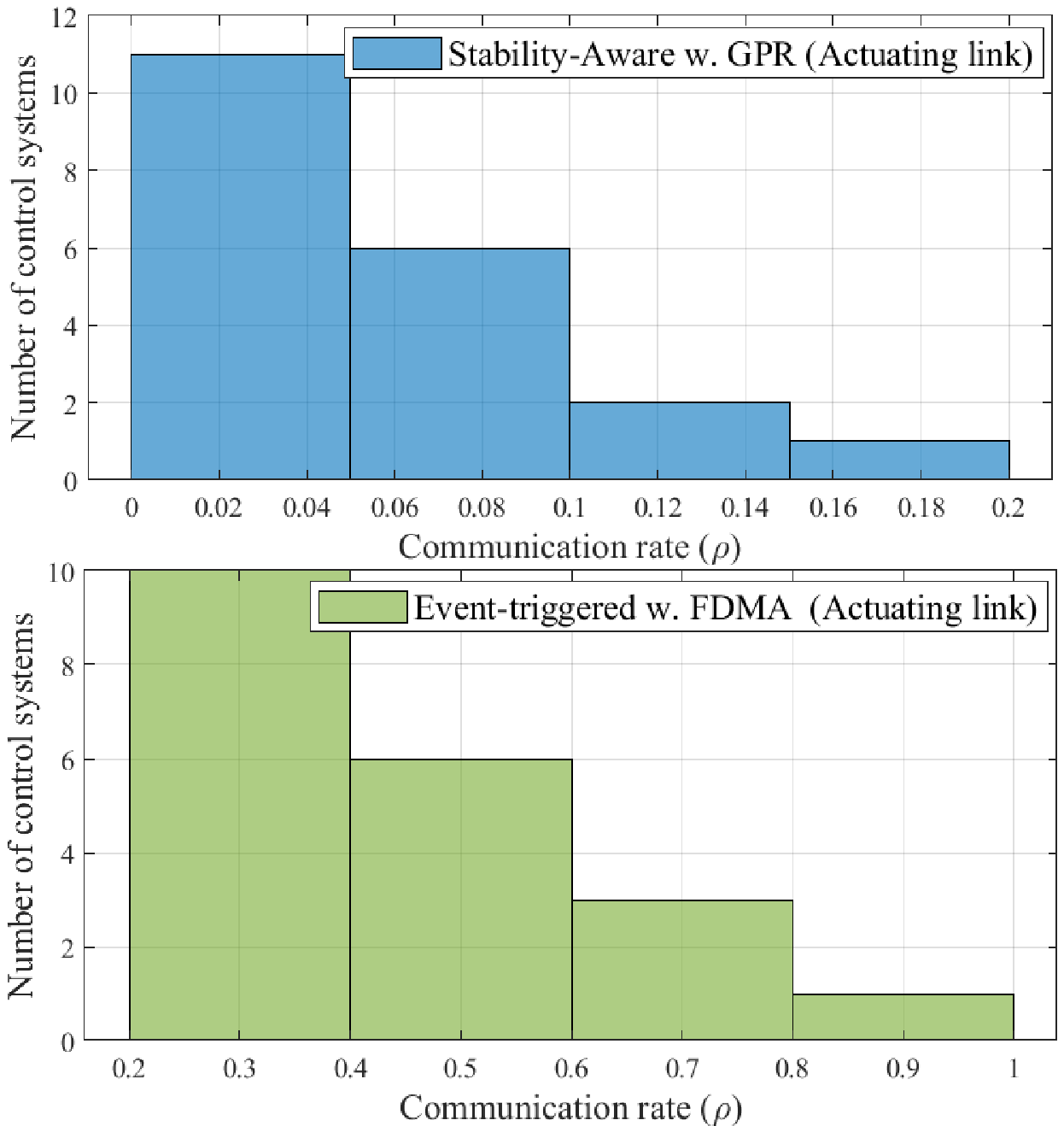} }
  \caption{Histogram of achieved communication rate in a large number of control system $\left( M=20 \right)$ of the proposed stability-aware and event-triggered with FDMA throughout $90$ control time steps.} \label{Fig21} 
\end{figure}

\noindent\textbf{Communication Rate Vs. Number of Control Systems}.\quad Fig.\ref{Fig21} presents a histogram of the achieved communication rates  for the sensing and actuating links for $M = 20$ during $90$ control time steps. The sensing/actuating communication rate is defined as the number of times the sensing/actuating link of a control system is scheduled divided by the time interval as $ n_{l} / K$. It is clear that the proposed approach achieves sensing communication rates concentrated in the range from $0$ to $0.4$ and actuating communication rates ranging from $0$ to $0.2$. On the other hand, the event-triggered with FDMA scheduling achieves wide sensing and actuating communication rates ranging from $0$ to $1$. The reason behind this result is that the stability-aware with GPR scheduling is adapting to both the channel and control states, and GPRs at the controller and actuator sides compensate for the missing received observations, hence improving the communication efficiency. Meanwhile, each control system, in the event-triggered with FDMA scheduling, only transmits its control state/action based on its control stability condition without taking into account channel states that result in increasing state/action estimation uncertainty from the adverse channel states, and in turn affecting control stability. Hence, each control system in the event-triggered with FDMA scheduling requires frequent transmissions to ensure control stability by applying appropriate action based on low communication uncertainty. Note that the range of the sensing communication rate is larger than that of the actuating communication rate since some control systems fail to transmit their states at the beginning affecting prediction credibility.  However, the number of control systems that require frequent scheduling in the actuating link is larger than that of the sensing link which stems from  the fast dynamics related to the inverted-pendulum system that require a quick appropriate action.

\begin{figure}[htb!]
  \centering
  \subfigure[Single control system out of  $M=2$.\label{Fig22.a}]{\includegraphics[width=.4\textwidth]{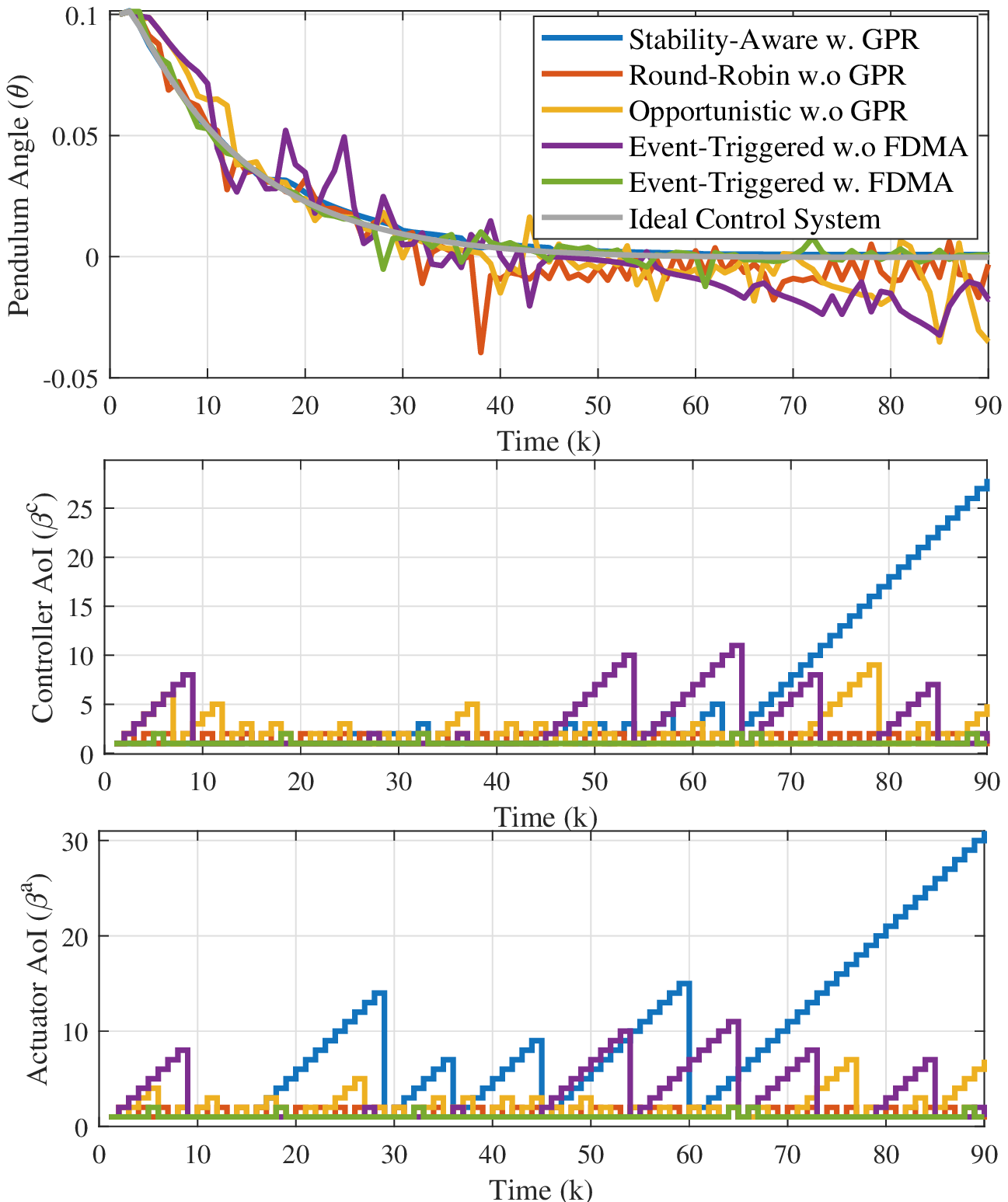}}
  \subfigure[Single control system out of $M=20$.\label{Fig22.b} ]{\includegraphics[width=.4\textwidth]{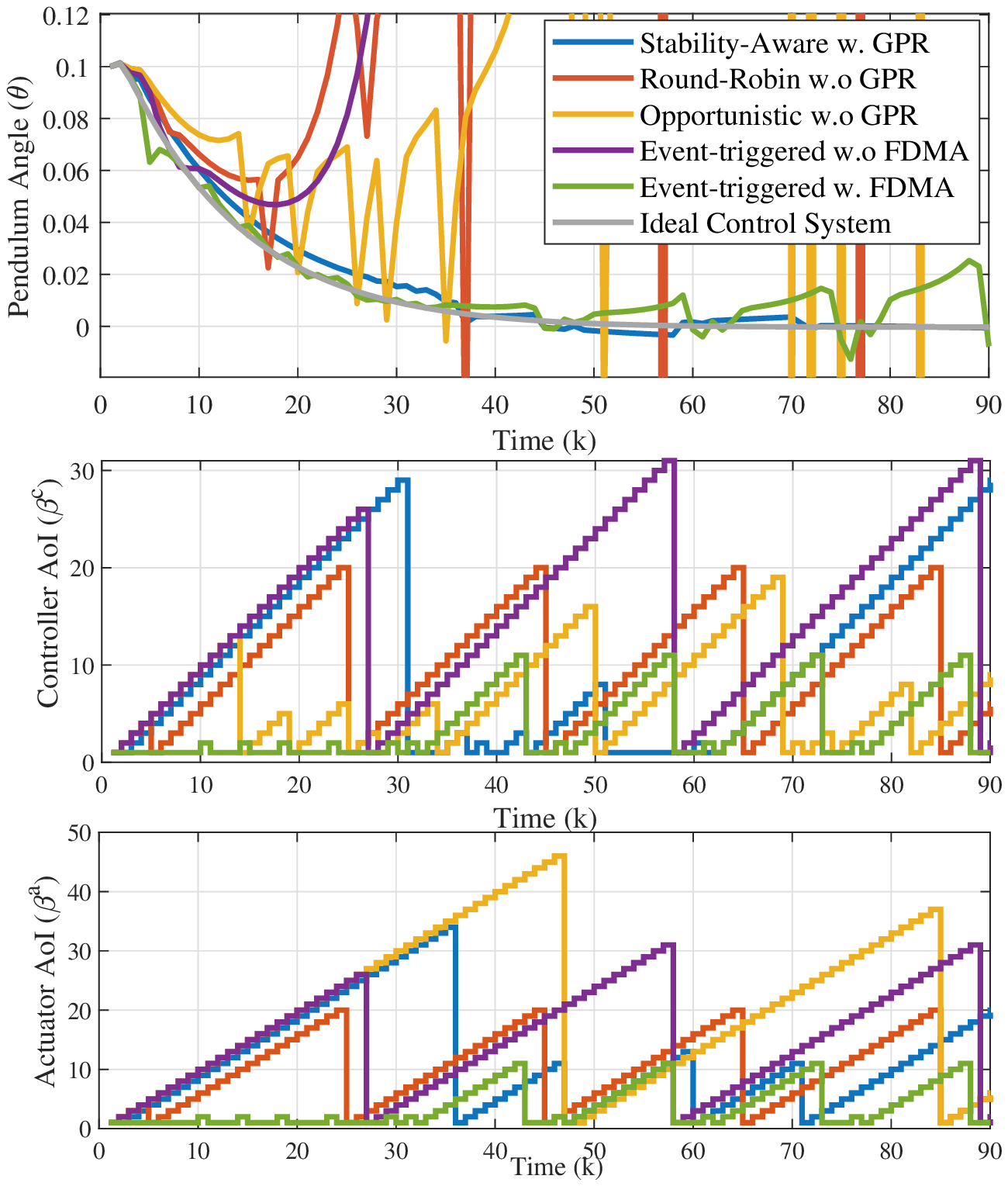}}
  \caption{Comparison of the state trajectory, controller AoI, and actuator AoI of a randomly chosen control system between the proposed stability-aware, round-robin, Opportunistic, event-triggered with and without FDMA, and ideal control system.} \label{Fig22}
\end{figure}

 \noindent\textbf{State Trajectory and Controller/Actuator AoI Vs. Time}.\quad To dive deeper into the benefits of the proposed stability-aware with GPR scheduling, we present in Fig.~\ref{Fig22} the state trajectory, controller AoI, and  actuator AoI of a randomly chosen control system in low and large number control systems regimes. In Fig~\ref{Fig22.a}, the state trajectory of the proposed stability-aware with GPR scheduling and event-triggered with FDMA scheduling, in a low number of control systems, are extremely close to that of the ideal control system where their pendulums remain upright over time. Meanwhile, the state trajectory of the opportunistic without GPR scheduling is slightly better than that of the event-triggered without FDMA scheduling by keeping all pendulums upright over time due to the scheduled control system with a favorable channel state. Finally, the state trajectory of the round-robin without GPR scheduling slightly matches the desired state at the cost of a high communication rate and fixed transmission power.
 
 As shown in Fig.~\ref{Fig22.a}, the controller and actuator AoI of the proposed stability-aware scheduling equals one, i.e., the sensing and actuating links of a control system are synchronously scheduled, until $15$ control time steps. This is to guarantee that the received actions to action GPR has a low state/action estimation uncertainty that affects the GPR state/action prediction stability and control stability. Then, the actuator AoI starts increasing compared to the controller AoI that remains at value one until $45$ control time steps, i.e., the sensing link of a control system is scheduled until $45$ control time steps while the actuating link is not scheduled. The rationale behind this result is to schedule either the sensing link or/and actuating link of a control system with favorable channel condition to ensure the state/action prediction credibility. Hence, the transmitted action, when the sensing link of a control system is not scheduled, depends on a credible predicted state. This shows the impact of the decoupled scheduling between the UL and DL communications compared to the coupled scheduling in terms of improving communication efficiency by reducing sensing and actuating communication rates while guaranteeing control stability.   For large numbers of control systems shown in Fig.~\ref{Fig22.b}, the control error of the  stability-aware with GPR scheduling and  event-triggered with FDMA scheduling  exponentially decay over time compared to the scheduling baselines. Meanwhile, the other scheduling baselines are exponentially growing over time due to the accumulated control error in the absence of appropriate action.

\begin{figure} [htb!]
  \centering
  \subfigure[Single control system out of $M=2$.\label{Fig13.a}]{\includegraphics[width=.4\linewidth]{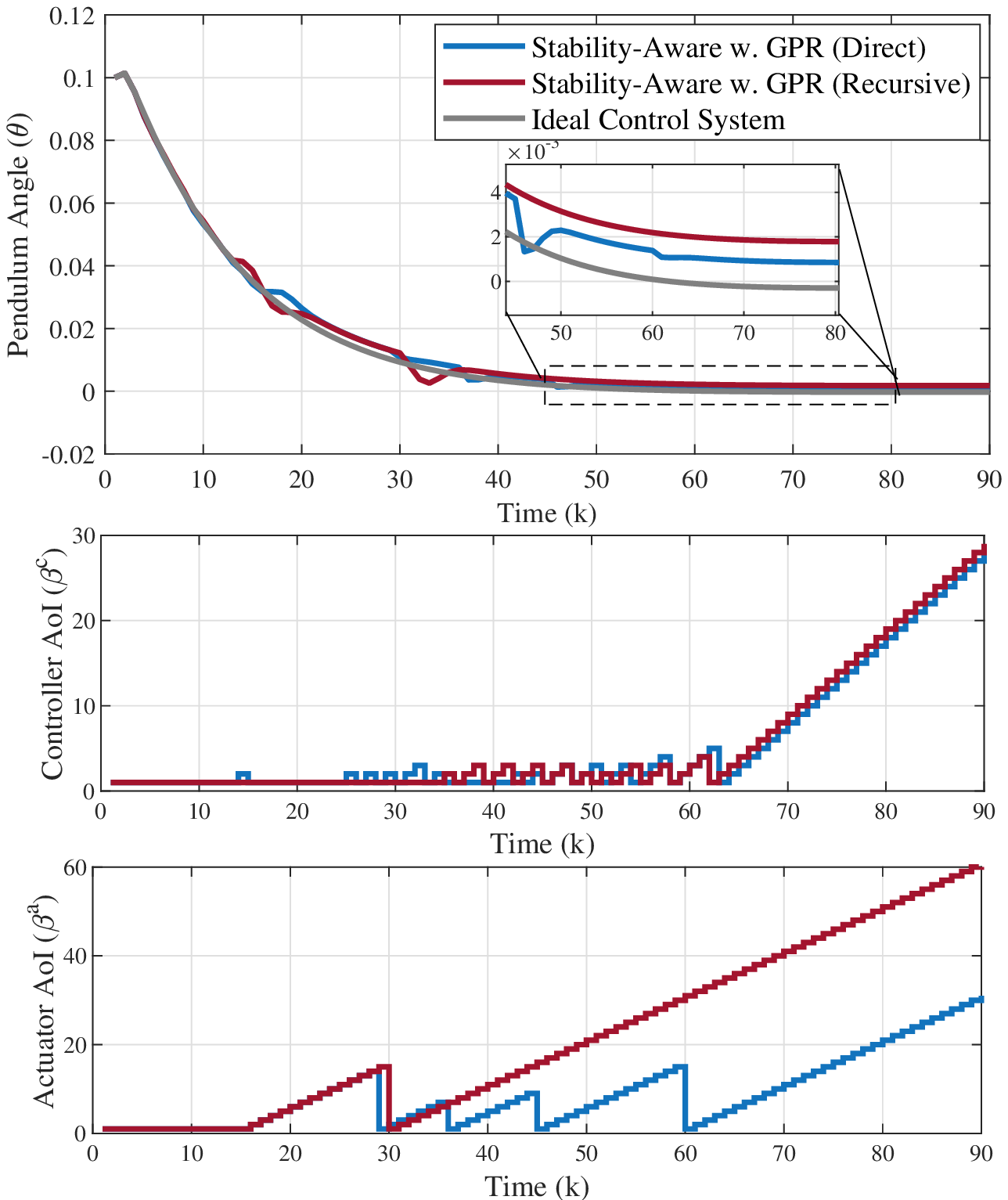}}
  \subfigure[Single control system out of  $M=20$.\label{Fig13.b}]{\includegraphics[width=.4\linewidth]{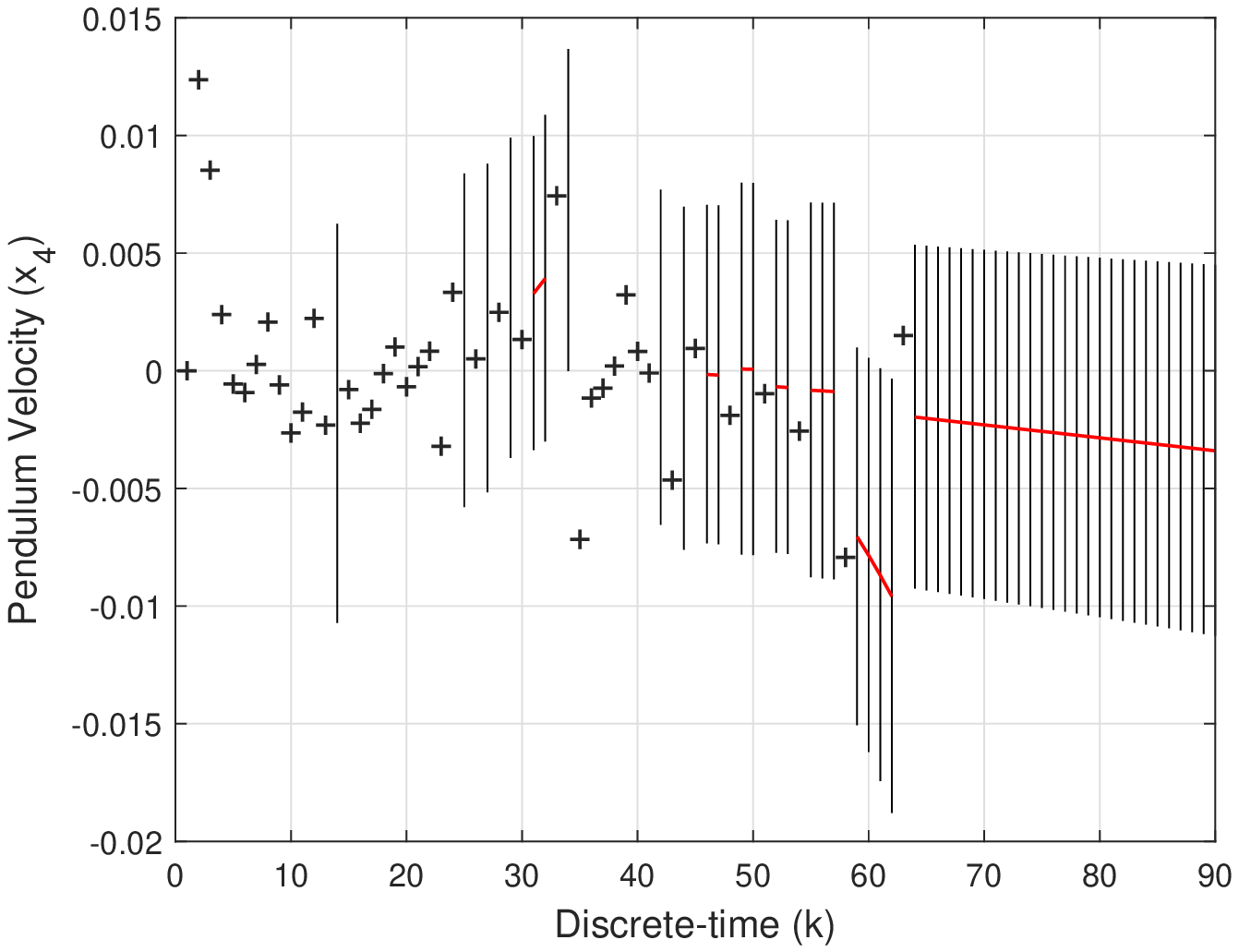}}
  \caption{Comparison of the state trajectory, controller AoI, and actuator AoI of a randomly chosen system between the proposed stability-aware w. GPR using both direct and recursive approaches.} \label{Fig13}
\end{figure}

 \noindent\textbf{ Time-Series Direct and Recursive GPR Approaches}.\quad Fig.~\ref{Fig13} illustrates the state trajectory, controller AoI, and actuator AoI of a randomly chosen control system in low and large number control systems and for two different time series GPR approaches. The recursive GPR approach uses the previous predicated and observed states/actions to predict the next control state/action, while the direct GPR approach only uses the previous observed states/actions to predict the next control state/action. It is interesting to observe that the state trajectory of the direct and recursive GPR approaches at the early phase are identical since the number of predicted states/actions in the training set of recursive GPR approach is smaller than the observed states/actions to predict future states/actions. However, when time increases, the state trajectory of the direct GPR approach outperforms the recursive GPR approach since the number of predicted states/actions in training set of the recursive GPR approach is greater than the observed states/actions. This in turn deteriorates the predication accuracy hindering control stability~\cite{sorjamaa2007methodology}.

 Moreover, the direct and recursive GPR approaches for a low number of control systems require frequent scheduling at the early phase (until $45$ control time steps in the sensing link), during which the amount of state observations ensures the state prediction credibility. Meanwhile, the recursive and direct GPR approaches for a low number of control systems require only $15$ control time steps in the actuating link, within which the amount of control actions ensures the action prediction stability. Fig.~\ref{Fig13} highlights that the AoI  increase deteriorates the control performance when increasing  prediction uncertainty until the training set has a sufficient number of observations to ensure steady-state state/action prediction. Finally, Fig.~\ref{Fig13} illustrates that the state/action prediction of the recursive GPR approach converges faster to the steady-state compared to the direct GPR approach. This affects the scheduling decision by increasing the controller and actuator AoIs of the recursive GPR approach at the cost of a low state/action prediction accuracy and control performance compared to the direct GPR approach.

\begin{figure}[htb!]
\centering
 \includegraphics[trim=2 2 20 20,clip, width=0.4\textwidth]{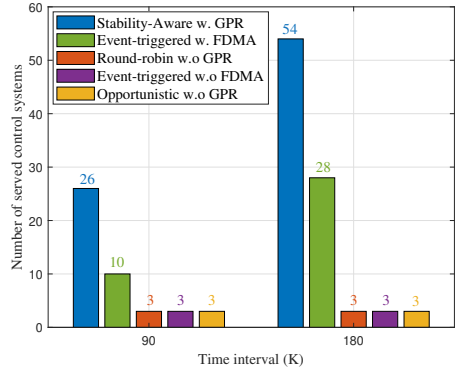} \\
\caption{ Total number of supported control systems using stability-aware with GPR, event-triggered with FDMA, round-robin, event-triggered without FDMA, and opportunistic scheduling.} \label{fig19}
 \end{figure}

 \noindent\textbf{ Number of Served Control Systems Vs. Time Interval}.\quad  Fig.~\ref{fig19} presents the final capacity of the scheduled control systems over two different time intervals. We assume that a scheduling method successfully control several controls systems for all control systems within $\| \theta_{i,k} \| \leq 0.05$ error region for $10$ independent simulation runs. As observed in Fig.~\ref{fig19}, the proposed approach has a significant impact compared to the baselines in terms of supporting a large number of control systems over time interval. The rationale behind this result is due to exploiting two GPRs at the controller and actuator sides and obtaining a sufficient number of observations in the GPR training sets increases. This in turn enhances the prediction credibility, communication efficiency, and control stability performance while supporting a large number of control systems.

\section{Conclusion}
\label{Conclusion}
In this work, we proposed a novel stability-aware scheduling algorithm based on the communication and control co-design exploiting analog uncoded communication and GPR based to reduce the required communication rate and ensure control stability through maintaining prediction stability. We performed extensive simulations for low-latency control systems to demonstrate the effectiveness of utilizing the GPR based approach, as well as, the effectiveness of decoupled scheduling between the UL and DL communications to support a large number of control systems compared to the scheduling baselines.
\appendices


\section{Proof of Lemma 1}

\label{App.lemma1}
Given the predicted state $\hat{\mathbf{x}}^{u}_{i,k}$ in~\eqref{eq9+a} and the state prediction error covariance matrix $\mathcal{J}^{u}_{i,k}$ in~\eqref{eq9+b}, it holds that the expected current Lyapunov value of the system is \vspace{-5pt}
{\small \begin{align} 
    \mathbb{E} \big[ \mathcal{L}(\mathbf{x}^{u}_{i,k}) \vert \hat{\mathbf{x}}^{u}_{i,k} \big]  & =  \mathbb{E} \big[ \mathbf{x}^{u^{T}}_{i,k} \, \mathcal{Z} \, \mathbf{x}^{u}_{i,k} \vert \hat{\mathbf{x}}^{u}_{i,k} \big]   = \mathbb{E} \big[ \left( \hat{\mathbf{x}}^{u}_{i,k} - \mathbf{e}^{u}_{i,k} \right)^{T} \mathcal{Z} \left( \hat{\mathbf{x}}^{u}_{i,k} - \mathbf{e}^{u}_{i,k} \right) \big]  \label{lemm1.1b} \\ & = \mathbb{E} \big[  \hat{\mathbf{x}}^{u^{T}}_{i,k} \mathcal{Z} \hat{\mathbf{x}}^{u}_{i,k} - \hat{\mathbf{x}}^{u^{T}}_{i,k} \mathcal{Z} \mathbf{e}^{u}_{i,k} - \mathbf{e}^{u^{T}}_{i,k} \mathcal{Z} \hat{\mathbf{x}}^{u}_{i,k} + \mathbf{e}^{u^{T}}_{i,k} \mathcal{Z} \mathbf{e}^{u}_{i,k}  \big], \qquad \quad \forall \; \mathbb{S}^{D}_{++}.  \label{lemm1.1c}
\end{align} } In~\eqref{lemm1.1c}, the first term is a constant as the expectation is taken w.r.t. the state prediction error $\mathbf{e}^{u}_{i,k}$, while the cross-terms such as $\mathbb{E} [ \hat{\mathbf{x}}^{u^{T}}_{i,k} \mathcal{Z} \mathbf{e}^{u}_{i,k}] $ and $ \mathbb{E} [\mathbf{e}^{u^{T}}_{i,k} \mathcal{Z} \hat{\mathbf{x}}^{u}_{i,k}] $ can be canceled since the predicted state $\hat{\mathbf{x}}^{u}_{i,k}$ and the state prediction error $\mathbf{e}^{u}_{i,k}$ are  uncorrelated. Hence, the expected current value of $\mathcal{L} (\mathbf{x}_{i,k})$ is given as \vspace{-5pt} \begin{equation} \small 
    \begin{aligned}
    \label{lemm1.2}
    \mathbb{E} \big[ \mathcal{L}(\mathbf{x}^{u}_{i,k}) \vert \hat{\mathbf{x}}^{u}_{i,k} \big] & = \hat{\mathbf{x}}^{u^{T}}_{i,k} \mathcal{Z} \hat{\mathbf{x}}^{u}_{i,k} + \text{Tr} \left[ \mathcal{Z} \; \mathbb{E} [ \mathbf{e}^{u^{T}}_{i,k} \mathbf{e}^{u}_{i,k} ] \right]  = \Vert \hat{\mathbf{x}}^{u}_{i,k} \Vert_{\mathcal{Z}^{\frac{1}{2}}}^{2} + \text{Tr} \left[ \mathcal{Z} \; \mathcal{J}^{u}_{i,k} \right],  \qquad  \qquad \forall \; \mathbb{S}^{D}_{++}
    \end{aligned}
\end{equation} where the last term in~\eqref{lemm1.2} is obtained via the expectation of the  quadratic form ${\scriptstyle \mathbb{E} [ \mathbf{e}^{u^{T}}_{i,k} \mathcal{Z} \mathbf{e}^{u}_{i,k} ]}$ w.r.t the state estimation error, i.e,  ${\scriptstyle \mathbb{E} [ \mathbf{e}^{u^{T}}_{i,k} \mathcal{Z} \mathbf{e}^{u}_{i,k} ] = ( \mathbb{E} [ \mathbf{e}^{u}_{i,k} ] )^{T}\mathcal{Z} ( \mathbb{E} [ \mathbf{e}^{u}_{i,k} ] ) + \text{Tr} [ \mathcal{Z} \; \mathcal{J}^{u}_{i,k} ]}$. It is observed in~\eqref{lemm1.2} that the expected current Lyapunov value grows larger as the predicted state $\hat{\mathbf{x}}^{u}_{i,k}$ is near instability and/or the prediction error covariance matrix  $\mathcal{J}^{u}_{i,k}$ is larger due to lack of sufficient observations. \hfill $\blacksquare$



\section{Proof of Lemma 2}
 \label{App.lemma2}
 As a result of the remote sensing-loop state evolution in~\eqref{eq22.2} and the open-loop state evolution in~\eqref{eq22.1}, the expected future value of $\mathcal{L} (\mathbf{x}^{u}_{i,k+1})$ in~\eqref{Future_Lyapunov} of the UL transmission is given as \vspace{-5pt}  \begin{equation} \small
    \begin{aligned}
    \label{lemm2.1}
    \mathbb{E} \left[ \mathcal{L} ( \mathbf{x}^{u}_{i,k+1}) \vert \hat{\mathbf{x}}^{u}_{i,k}, \hat{\mathbf{u}}^{d}_{i,k}, \mathbf{H}^{u}_{i,k},  P^{u}_{i,k}  \right] = & \xi^{u}_{i,k} \; \mathbb{E} \left[ \mathcal{L} (\mathbf{x}^{s}_{i,k+1}) \vert \hat{\mathbf{x}}^{u}_{i,k}, \hat{\mathbf{u}}^{d}_{i,k}, \mathbf{H}^{u}_{i,k}, P^{u}_{i,k} \right] \\ & + \left( 1 - \xi^{u}_{i,k} \right) \; \mathbb{E} \left[ \mathcal{L} (\mathbf{x}^{o}_{i,k+1}) \vert \hat{\mathbf{x}}^{u}_{i,k}, \hat{\mathbf{u}}^{d}_{i,k}, \mathbf{H}^{u}_{i,k}, P^{u}_{i,k} \right],
    \end{aligned}
 \end{equation} For a given predicted state $\hat{\mathbf{x}}^{u}_{i,k}$, state prediction error covariance matrix $\mathcal{J}^{u}_{i,k}$, predicted action $\hat{\mathbf{u}}^{d}_{i,k}$, action prediction error covariance matrix $\mathcal{J}^{d}_{i,k}$, wireless UL channel $\mathbf{H}^{u}_{i,k}$, and UL transmission power $P^{u}_{i,k}$, the expected future value of $\mathcal{L} ( \mathbf{x}^{u}_{i,k+1})$ of the UL transmission in~\eqref{lemm2.1} is given as \vspace{-3pt} \begin{equation} \small
        \begin{aligned}  
         \label{lemm2.2}
        \mathbb{E} & \left[ \mathcal{L} (\mathbf{x}^{u}_{i,k+1}) \vert \hat{\mathbf{x}}^{u}_{i,k}, \hat{\mathbf{u}}^{d}_{i,k}, \mathbf{H}^{u}_{i,k},  P^{u}_{i,k}  \right]   \\ &   \; \overset{(1)}{=} \xi^{u}_{i,k} \mathbb{E} \left[ \mathbf{x}^{s^{T}}_{i,k+1} \mathcal{Z} \mathbf{x}^{s}_{i,k+1} \vert \hat{\mathbf{x}}^{u}_{i,k}, \hat{\mathbf{u}}^{d}_{i,k}, \mathbf{H}^{u}_{i,k}, P^{u}_{i,k} \right] + \left( 1 - \xi^{u}_{i,k} \right) \; \mathbb{E} \left[ \mathbf{x}^{o^{T}}_{i,k+1} \mathcal{Z} \mathbf{x}^{o}_{i,k+1} \vert \hat{\mathbf{x}}^{u}_{i,k}, \hat{\mathbf{u}}^{d}_{i,k}, \mathbf{H}^{u}_{i,k}, P^{u}_{i,k}   \right] \\ & \overset{(2)}{=} \xi^{u}_{i,k} \mathbb{E} \left[ \left( \mathbf{A}^{c}_{i} \hat{\mathbf{x}}^{u}_{i,k} - \mathbf{A}^{c}_{i} \mathbf{e}^{u}_{i,k} -  \mathbf{B}_{i} \mathbf{\Phi}_{i} \mathbf{v}^{u}_{i,k} - \mathbf{B}_{i} \mathbf{e}^{d}_{i,k} + \mathbf{w}_{k} \right)^{T} \mathcal{Z} \left( \mathbf{A}^{c}_{i} \hat{\mathbf{x}}^{u}_{i,k} - \mathbf{A}^{c}_{i} \mathbf{e}^{u}_{i,k} -  \mathbf{B}_{i} \mathbf{\Phi}_{i} \mathbf{v}^{u}_{i,k} - \mathbf{B}_{i} \mathbf{e}^{d}_{i,k} + \mathbf{w}_{k} \right)  \right] \\ & + (1- \xi^{u}_{i,k}) \mathbb{E} \left[ \left( \mathbf{A}^{c}_{i} \hat{\mathbf{x}}^{u}_{i,k} - \mathbf{A}^{c}_{i} \mathbf{e}^{u}_{i,k} -  \mathbf{B}_{i} \mathbf{\Phi}_{i} \mathbf{e}^{u}_{i,k} - \mathbf{B}_{i} \mathbf{e}^{d}_{i,k} + \mathbf{w}_{k} \right)^{T} \mathcal{Z} \left( \mathbf{A}^{c}_{i} \hat{\mathbf{x}}^{u}_{i,k} - \mathbf{A}^{c}_{i} \mathbf{e}^{u}_{i,k} -  \mathbf{B}_{i} \mathbf{\Phi}_{i} \mathbf{e}^{u}_{i,k} - \mathbf{B}_{i} \mathbf{e}^{d}_{i,k} + \mathbf{w}_{k} \right)  \right] \\ & \overset{(3)}{=} \xi^{u}_{i,k} \left[ \Vert \mathbf{A}^{c}_{i} \hat{\mathbf{x}}^{u}_{i,k} \Vert_{\mathcal{Z}^{\frac{1}{2}}}^{2} + \text{Tr} \left[ \mathbf{A}^{c^{T}}_{i} \mathcal{Z} \mathbf{A}^{c}_{i} \mathcal{J}^{u}_{i,k} \right]  + \text{Tr} \left[ \left( \mathbf{B}_{i} \mathbf{\Phi}_{i} \right)^{T} \mathcal{Z}  \left( \mathbf{B}_{i} \mathbf{\Phi}_{i} \right)\mathbf{V}^{u}_{i,k} \right] + \text{Tr} \left[ \mathbf{B}^{T}_{i} \mathcal{Z} \mathbf{B}_{i} \mathcal{J}^{d}_{i,k} \right] + \text{Tr} \left[ \mathcal{Z} \mathbf{W} \right] \right] \\ & +  ( 1 - \xi^{u}_{i,k} )  
        \left[ \Vert \mathbf{A}^{c}_{i} \hat{\mathbf{x}}^{u}_{i,k} \Vert_{\mathcal{Z}^{\frac{1}{2}}}^{2} + \text{Tr} \left[ \mathbf{A}^{c^{T}}_{i} \mathcal{Z} \mathbf{A}^{c}_{i} \mathcal{J}^{u}_{i,k} \right]  + \text{Tr} \left[ \left( \mathbf{B}_{i} \mathbf{\Phi}_{i} \right)^{T} \mathcal{Z}  \left( \mathbf{B}_{i} \mathbf{\Phi}_{i} \right)\mathcal{J}^{u}_{i,k} \right] + \text{Tr} \left[ \mathbf{B}^{T}_{i} \mathcal{Z} \mathbf{B}_{i} \mathcal{J}^{d}_{i,k} \right] + \text{Tr} \left[ \mathcal{Z} \mathbf{W} \right] \right].
        \end{aligned}
    \end{equation} Step $(1)$ is a result of using the quadratic Lyapunov function. The step $(2)$ holds when applying the remote sensing-loop and open-loop state evolution in~\eqref{eq22.2} and~\eqref{eq22.1}, respectively. The step $(3)$ holds using the expectation in~\eqref{lemm2.2} with respect to the state estimation error $\mathbf{e}^{u}_{i,k}$, the action prediction error $\mathbf{e}^{d}_{i,k}$, and the plant noise $\mathbf{w}_{k}$. As a consequence of obtaining the expected current Lyapunov value in~\eqref{lemm1.2} and the expected future Lyapunov value of the UL transmission in~\eqref{lemm2.2}, the control stability constraint in~\eqref{Future_Lyapunov} for the UL transmission is \vspace{-4pt}  \begin{equation} \small
    \begin{aligned}
    \label{lemm2.3}
    & \xi^{u}_{i,k}  \mathbb{E} \left[ \mathcal{L} (\mathbf{x}^{s}_{i,k+1}) \vert \hat{\mathbf{x}}^{u}_{i,k}, \hat{\mathbf{u}}^{d}_{i,k}, \mathbf{H}^{u}_{i,k}, P^{u}_{i,k} \right] + ( 1-  \xi^{u}_{i,k}) \mathbb{E} \left[ \mathcal{L} (\mathbf{x}^{o}_{i,k+1}) \vert \hat{\mathbf{x}}^{u}_{i,k}, \hat{\mathbf{u}}^{d}_{i,k}, \mathbf{H}^{u}_{i,k}, P^{u}_{i,k} \right] \leq \zeta_{i} \mathbb{E} \left[ \mathcal{L} (\mathbf{x}^{u}_{i,k}) \vert \hat{\mathbf{x}}^{u}_{i,k} \right] \\ & = \xi^{u}_{i,k} \left[ \text{Tr} \left[  ( \mathbf{B}_{i} \mathbf{\Phi}_{i} )^{T} \mathcal{Z} (\mathbf{B}_{i} \mathbf{\Phi}_{i} ) \mathbf{V}^{u}_{i,k} \right] - \text{Tr} \left[  ( \mathbf{B}_{i} \mathbf{\Phi}_{i} )^{T} \mathcal{Z} (\mathbf{B}_{i} \mathbf{\Phi}_{i} ) \mathcal{J}^{u}_{i,k} \right] \right] + \Vert \mathbf{A}^{c}_{i} \hat{\mathbf{x}}^{u}_{i,k} \Vert_{\mathcal{Z}^{\frac{1}{2}}}^{2} + \text{Tr} \left[ \mathbf{A}^{T}_{i} \mathcal{Z} \mathbf{A}_{i} \mathcal{J}^{u}_{i,k} \right] \\ & \qquad + \text{Tr} \left[ \mathbf{B}^{T}_{i} \mathcal{Z} \mathbf{B}_{i} \mathcal{J}^{d}_{i,k} \right] + \text{Tr} \left[ \mathcal{Z} \mathbf{W} \right] \leq \zeta_{i} \Vert \hat{\mathbf{x}}^{u}_{i,k} \Vert_{\mathcal{Z}^{\frac{1}{2}}}^{2} + \zeta_{i} \text{Tr} \left[ \mathcal{Z} \mathcal{J}^{u}_{i,k} \right] 
    \end{aligned}
\end{equation}  After rearranging the terms in~\eqref{lemm2.3}, the constraint on the UL transmission indicator variable is \vspace{-3pt} { \small \begin{align} 
 \label{lemm2.4}
 \xi^{u}_{i,k} \geq  \frac{  \Vert \left( \mathbf{A}^{c}_{i} - \zeta_{i} \mathbf{I}_{D}  \right) \hat{\mathbf{x}}^{u}_{i,k} \Vert_{\mathcal{Z}^{\frac{1}{2}}}^{2} +  \text{Tr} \left[ \left( \mathbf{A}^{T}_{i} \mathcal{Z} \mathbf{A}_{i}  - \zeta_{i}  \mathcal{Z} \right) \mathcal{J}^{u}_{i,k} \right]   + \text{Tr} \left[ \mathbf{B}^{T}_{i} \mathcal{Z} \mathbf{B}_{i} \mathcal{J}^{d}_{i,k} \right] + \text{Tr} \left[ \mathcal{Z} \mathbf{W} \right] }{ \text{Tr} \left[  \left( \mathbf{B}_{i} \mathbf{\Phi}_{i} \right)^{T} \mathcal{Z}  \left( \mathbf{B}_{i} \mathbf{\Phi}_{i} \right) \mathcal{J}^{u}_{i,k} \right] -  \text{Tr} \left[  \left( \mathbf{B}_{i} \mathbf{\Phi}_{i} \right)^{T} \mathcal{Z}  \left( \mathbf{B}_{i} \mathbf{\Phi}_{i} \right) \mathbf{V}^{u}_{i,k} \right] }. \end{align} }  To capture the overall state evolution of each control system in the UL  over time interval of length $K$, according to the time-averaged Lyapunov~\cite{michel2009stability}, (53) is summed over time ${ \scriptstyle k \in \{0,\cdots,K-1 \}}$, then the result is divided by $K$ and taking the limit as time tends to infinity. This yields the UL transmission indicator variable constraint in~\eqref{eq32}.  
 
 Similarly, we obtain the DL transmission indicator variable constraint in~\eqref{eq33} according to the remote actuating-loop state evolution in~\eqref{eq22.3} and  open-loop state evolution in~\eqref{eq22.1}. Finally, the expected future Lyapunov value of the UL-DL coupled transmission is obtained using the closed-loop state evolution in~\eqref{eq22.4} and the open-loop state evolution in~\eqref{eq22.1} where \vspace{-3pt} \begin{equation} \small 
    \begin{aligned}
    \label{lemm2.9}
    \mathbb{E} \Big[ \mathcal{L} (\mathbf{x}^{u}_{i,k+1}) \vert \hat{\mathbf{x}}^{u}_{i,k}, \hat{\mathbf{u}}^{d}_{i,k},  \mathbf{H}^{u}_{i,k}, P^{u}_{i,k},  &\mathbf{H}^{d}_{i,k},  P^{d}_{i,k}  \Big]  =  \xi^{u}_{i,k}\;\xi^{d}_{i,k} \; \mathbb{E} \left[ \mathcal{L} (\mathbf{x}^{c}_{i,k+1}) \vert \hat{\mathbf{x}}^{u}_{i,k}, \hat{\mathbf{u}}^{d}_{i,k}, \mathbf{H}^{u}_{i,k}, P^{u}_{i,k}, \mathbf{H}^{d}_{i,k}, P^{d}_{i,k} \right] \\ & + \left( 1 - \xi^{u}_{i,k}\;\xi^{d}_{i,k} \right) \; \mathbb{E} \left[ \mathcal{L} (\mathbf{x}^{o}_{i,k+1}) \vert \hat{\mathbf{x}}^{u}_{i,k}, \hat{\mathbf{u}}^{d}_{i,k}, \mathbf{H}^{u}_{i,k}, P^{u}_{i,k}, \mathbf{H}^{d}_{i,k}, P^{d}_{i,k} \right],
    \end{aligned}
\end{equation} For a given predicted state $\hat{\mathbf{x}}^{u}_{i,k}$, the state prediction error covariance matrix $\mathcal{J}^{u}_{i,k}$, the predicted action $\hat{\mathbf{u}}^{d}_{i,k}$, the action prediction error covariance matrix $\mathcal{J}^{d}_{i,k}$, the wireless UL channel $\mathbf{H}^{u}_{i,k}$, the wireless DL channel $\mathbf{H}^{d}_{i,k}$, the UL transmission power $P^{u}_{i,k}$, and the DL transmission power $P^{d}_{i,k}$, the expected future value of $\mathcal{L} ( \mathbf{x}^{u}_{i,k+1})$ of the UL-DL coupling transmission in~\eqref{lemm2.9} is given as 
\begin{equation} 
 \small        \begin{aligned}  
         \label{lemm2.10} 
        \mathbb{E} & \left[ \mathcal{L} (\mathbf{x}^{u}_{i,k+1}) \vert \hat{\mathbf{x}}^{u}_{i,k}, \hat{\mathbf{u}}^{d}_{i,k}, \mathbf{H}^{u}_{i,k},  P^{u}_{i,k}, \mathbf{H}^{d}_{i,k},  P^{d}_{i,k}  \right] =   \xi^{u}_{i,k} \; \xi^{d}_{i,k} \mathbb{E} \left[ \mathbf{x}^{c^{T}}_{i,k+1} \mathcal{Z} \mathbf{x}^{c}_{i,k+1} \vert \hat{\mathbf{x}}^{u}_{i,k}, \hat{\mathbf{u}}^{d}_{i,k}, \mathbf{H}^{u}_{i,k}, P^{u}_{i,k}, \mathbf{H}^{d}_{i,k}, P^{d}_{i,k} \right] \\& \hspace{160pt} + \left( 1 - \xi^{u}_{i,k} \; \xi^{d}_{i,k} \right) \; \mathbb{E} \left[ \mathbf{x}^{o^{T}}_{i,k+1} \mathcal{Z} \mathbf{x}^{o}_{i,k+1} \vert \hat{\mathbf{x}}^{u}_{i,k}, \hat{\mathbf{u}}^{d}_{i,k}, \mathbf{H}^{u}_{i,k}, P^{u}_{i,k}  \mathbf{H}^{d}_{i,k}, P^{d}_{i,k}   \right] \\ & = \xi^{u}_{i,k} \; \xi^{d}_{i,k} \mathbb{E} \left[ \left( \mathbf{A}^{c}_{i} \hat{\mathbf{x}}^{u}_{i,k} - \mathbf{A}^{c}_{i} \mathbf{e}^{u}_{i,k} - \mathbf{B}_{i} \mathbf{\Phi}_{i}  \mathbf{v}^{c}_{i,k} -  \mathbf{B}_{i}  \mathbf{v}^{d}_{i,k}  + \mathbf{w}_{k} \right)^{T} \mathcal{Z} \left( \mathbf{A}^{c}_{i} \hat{\mathbf{x}}^{u}_{i,k} - \mathbf{A}^{c}_{i} \mathbf{e}^{u}_{i,k} - \mathbf{B}_{i} \mathbf{\Phi}_{i}  \mathbf{v}^{c}_{i,k} -  \mathbf{B}_{i}  \mathbf{v}^{d}_{i,k}  + \mathbf{w}_{k} \right)  \right] \\ & + (1- \xi^{u}_{i,k} \; \xi^{d}_{i,k}) \mathbb{E} \left[ \left( \mathbf{A}^{c}_{i} \hat{\mathbf{x}}^{u}_{i,k} - \mathbf{A}^{c}_{i} \mathbf{e}^{u}_{i,k} -  \mathbf{B}_{i} \mathbf{\Phi}_{i} \mathbf{e}^{u}_{i,k} - \mathbf{B}_{i} \mathbf{e}^{d}_{i,k} + \mathbf{w}_{k} \right)^{T} \mathcal{Z}  \left( \mathbf{A}^{c}_{i} \hat{\mathbf{x}}^{u}_{i,k} - \mathbf{A}^{c}_{i} \mathbf{e}^{u}_{i,k} -  \mathbf{B}_{i} \mathbf{\Phi}_{i} \mathbf{e}^{u}_{i,k} - \mathbf{B}_{i} \mathbf{e}^{d}_{i,k} + \mathbf{w}_{k} \right)  \right] \\ & 
        = \xi^{u}_{i,k} \; \xi^{d}_{i,k} \left[ \Vert \mathbf{A}^{c}_{i} \hat{\mathbf{x}}^{u}_{i,k} \Vert_{\mathcal{Z}^{\frac{1}{2}}}^{2} + \text{Tr} \left[ \mathbf{A}^{c^{T}}_{i} \mathcal{Z} \mathbf{A}^{c}_{i} \mathcal{J}^{u}_{i,k} \right] + \text{Tr} \left[ (\mathbf{B}_{i} \mathbf{\Phi}_{i})^{T} \mathcal{Z} (\mathbf{B}_{i} \mathbf{\Phi}_{i}) \mathbf{V}^{u}_{i,k} \right]
        + \text{Tr} \left[ \mathbf{B}^{T}_{i} \mathcal{Z} \mathbf{B}_{i} \mathbf{V}^{d}_{i,k} \right] + \text{Tr} \left[ \mathcal{Z} \mathbf{W} \right] \right] \\ & +  ( 1 - \xi^{u}_{i,k} \; \xi^{d}_{i,k} )  
        \left[ \Vert \mathbf{A}^{c}_{i} \hat{\mathbf{x}}^{u}_{i,k} \Vert_{\mathcal{Z}^{\frac{1}{2}}}^{2} + \text{Tr} \left[ \mathbf{A}^{c^{T}}_{i} \mathcal{Z} \mathbf{A}^{c}_{i} \mathcal{J}^{u}_{i,k} \right] + \text{Tr} \left[ (\mathbf{B}^{T}_{i} \mathbf{\Phi}_{i})^{T} \mathcal{Z} (\mathbf{B}^{T}_{i} \mathbf{\Phi}_{i}) \mathcal{J}^{u}_{i,k}  \right]  + \text{Tr} \left[ \mathbf{B}^{T}_{i} \mathcal{Z} \mathbf{B}_{i} \mathcal{J}^{d}_{i,k} \right] + \text{Tr} \left[ \mathcal{Z} \mathbf{W} \right] \right]
        \end{aligned}
    \end{equation} Given the expected current value of $\mathcal{L} (\mathbf{x}^{u}_{i,k})$ in~\eqref{lemm1.2} and  future of $\mathcal{L} (\mathbf{x}^{u}_{i,k+1})$ of the UL-DL coupled transmission in~\eqref{lemm2.10}, the control stability constraint in~\eqref{Future_Lyapunov} for  UL-DL coupled transmission is   \begin{equation} \small
    \begin{aligned}
    \label{lemm2.11}
    & \xi^{u}_{i,k} \; \xi^{d}_{i,k}  \mathbb{E} \left[ \mathcal{L} (\mathbf{x}^{a}_{i,k+1}) \vert \hat{\mathbf{x}}^{u}_{i,k}, \hat{\mathbf{u}}^{d}_{i,k}, \mathbf{H}^{u}_{i,k}, P^{u}_{i,k} \mathbf{H}^{d}_{i,k}, P^{d}_{i,k} \right]  + ( 1-  \xi^{u}_{i,k} \; \xi^{d}_{i,k} )  \mathbb{E} \Big[ \mathcal{L} (\mathbf{x}^{o}_{i,k+1}) \vert \hat{\mathbf{x}}^{u}_{i,k}, \hat{\mathbf{u}}^{d}_{i,k}, \mathbf{H}^{u}_{i,k}, P^{u}_{i,k}  \mathbf{H}^{d}_{i,k}, P^{d}_{i,k} \Big] \\& \leq \zeta_{i} \mathbb{E} \left[ \mathcal{L} (\mathbf{x}^{u}_{i,k}) \vert \hat{\mathbf{x}}^{u}_{i,k} \right]  = \xi^{u}_{i,k} \; \xi^{d}_{i,k}  \Big[ \text{Tr} \left[ (\mathbf{B}_{i} \mathbf{\Phi}_{i} )^{T} \mathcal{Z} (\mathbf{B}_{i} \mathbf{\Phi}_{i} )  \mathbf{V}^{u}_{i,k} \right] + \text{Tr} \left[  \mathbf{B}_{i}^{T} \mathcal{Z} \mathbf{B}_{i}  \mathbf{V}^{d}_{i,k} \right] - \text{Tr} \left[ (\mathbf{B}_{i} \mathbf{\Phi}_{i})^{T} \mathcal{Z} (\mathbf{B}_{i} \mathbf{\Phi}_{i}) \mathcal{J}^{u}_{i,k} \right] \\ & - \text{Tr} \left[ \mathbf{B}_{i}^{T} \mathcal{Z} \mathbf{B}_{i}  \mathcal{J}^{d}_{i,k} \right] \Big] + \Vert \mathbf{A}^{c}_{i} \hat{\mathbf{x}}^{u}_{i,k} \Vert_{\mathcal{Z}^{\frac{1}{2}}}^{2} + \text{Tr} \left[ \mathbf{A}^{T}_{i} \mathcal{Z} \mathbf{A}_{i} \mathcal{J}^{u}_{i,k} \right] + \text{Tr} \left[ \mathbf{B}^{T}_{i} \mathcal{Z} \mathbf{B}_{i} \mathcal{J}^{d}_{i,k} \right] + \text{Tr} \left[ \mathcal{Z} \mathbf{W} \right] \leq \zeta_{i} \Vert \hat{\mathbf{x}}^{u}_{i,k} \Vert_{\mathcal{Z}^{\frac{1}{2}}}^{2} + \zeta_{i} \text{Tr} \left[ \mathcal{Z} \mathcal{J}^{u}_{i,k} \right] 
    \end{aligned}
\end{equation} The  UL-DL transmission indicator variable constraint after rearranging the terms in~\eqref{lemm2.11} is:  { \small \begin{align} 
 \label{lemm2.12} 
  \xi^{u}_{i,k} \; \xi^{d}_{i,k} \geq  \frac{ 
 \Vert \left( \mathbf{A}^{c}_{i} - \zeta_{i} \mathbf{I}_{D}  \right) \hat{\mathbf{x}}^{u}_{i,k} \Vert_{\mathcal{Z}^{\frac{1}{2}}}^{2} +  \text{Tr} \left[ \left( \mathbf{A}^{T}_{i} \mathcal{Z} \mathbf{A}_{i}  - \zeta_{i}  \mathcal{Z} \right) \mathcal{J}^{u}_{i,k} \right]   + \text{Tr} \left[ \mathbf{B}^{T}_{i} \mathcal{Z} \mathbf{B}_{i} \mathcal{J}^{d}_{i,k} \right] + \text{Tr} \left[ \mathcal{Z} \mathbf{W} \right]}{ 
 \text{Tr} \left[ \left( (\mathbf{B}_{i} \mathbf{\Phi}_{i})^{T} \mathcal{Z}  (\mathbf{B}_{i} \mathbf{\Phi}_{i} ) \right) (\mathcal{J}^{u}_{i,k} - \mathbf{V}^{u}_{i,k})  \right]  + \text{Tr} \left[ \mathbf{B}_{i}^{T} \mathcal{Z}  \mathbf{B}_{i}  (\mathcal{J}^{d}_{i,k} - \mathbf{V}^{d}_{i,k}   \right] }
\end{align} } At  last, we apply the time-averaged Lyapunov~\cite{michel2009stability} to obtain the overall state evolution per control system in the UL-DL coupled transmission and  UL-DL coupled transmission indicator variable constraint in~\eqref{eq33e}. \hfill $\blacksquare$

\nocite{*}
\bibliographystyle{IEEEtran}
\bibliography{IEEEabrv,mybibfile}

\end{document}